\documentclass[longauth]{aa} 
\usepackage{graphicx}
\usepackage{txfonts}
\usepackage{color}
\usepackage{ulem}
\usepackage{natbib}
\bibpunct{(}{)}{;}{a}{}{,} 

\def\ie{{\it i.e.\ }}

\def\del{\delta}              
\def\hpc{$h^{-1}$Mpc }

\def\z{$\,${\it z}$\,$}

\def\cmos{C\textsc{mocks }}
\def\omos{O\textsc{mocks }}
\def\cmo{C\textsc{mocks}}
\def\omo{O\textsc{mocks}}
\def\ucmos{C\textsc{mock }}
\def\uomos{O\textsc{mock }}
\def\ucmo{C\textsc{mock}}
\def\uomo{O\textsc{mock}}

\begin{document}
   \title{Comparison of the VIMOS-VLT Deep Survey with the Munich
   semi-analytical model.} 

   \subtitle{II. The colour-density relation up to $z\sim 1.5$}

  \titlerunning{The VVDS colour-density relation at $z\leq 1.5$: model vs observations.}

  \author{O.~Cucciati\inst{1}
          \and
          G.~De Lucia\inst{1}
          \and
          E.~Zucca\inst{2}
          \and
          A.~Iovino\inst{3}
          \and
          S.~de la Torre\inst{4}
          \and
          L.~Pozzetti\inst{2}
          \and
	  J.~Blaizot\inst{5}
	  \and
          G.~Zamorani\inst{2}
          \and
          M.~Bolzonella\inst{2}
          \and
          D.~Vergani\inst{6}
	  \and 
	  S.~Bardelli\inst{2}
	  \and
	  L.~Tresse\inst{7}	
          \and
          A.~Pollo\inst{8,9}
          }

   \offprints{Olga Cucciati (cucciati@oats.inaf.it)}

   \institute{INAF-Osservatorio Astronomico di Trieste - Via Tiepolo 11, I-34143  Trieste, Italy
         \and INAF-Osservatorio Astronomico di Bologna, via Ranzani 1, 40127 Bologna, Italy
         \and INAF-Osservatorio Astronomico di Brera, Via Brera 28, I-20021, Milan, Italy
         \and SUPA, Institute for Astronomy, University of Edinburgh, Royal Observatory, Blackford Hill, EH9 3HJ Edinburgh, UK
	 \and Centre de Recherche Astrophysique de Lyon, UMR 5574, Universit\'e Claude Bernard Lyon-\'Ecole Normale Sup\'erieure de Lyon-CNRS, 69230 Saint-Genis Laval, France
         \and INAF-IASFBO, Via P.~Gobetti 101, I-40129, Bologna, Italy 
	 \and Aix Marseille Universit\'e, CNRS, LAM (Laboratoire d'Astrophysique de Marseille) UMR 7326, 13388, Marseille, France
         \and The Andrzej Soltan Institute for Nuclear Studies, ul. Hoza 69, 00-681 Warszawa, Poland
         \and Astronomical Observatory of the Jagiellonian University, ul. Orla 171, PL-30-244, Krak{\'o}w, Pola
}

 
  \abstract
    {}
   {Our aim is to perform on galaxy mock catalogues the
   same colour-density analysis carried out by Cucciati et al. (2006)
   on a 5$h^{-1}$Mpc scale using the VIMOS-VLT Deep Survey
   (VVDS), and to compare the results from mocks
   with observed data. This allows us  to test galaxy evolution
    in mocks, and to understand the relation between the studied
    environment and the underlying dark matter distribution.}
   {We used galaxy mock catalogues, with the same flux limits as
   the VVDS-Deep ($I_{AB} \leq 24$) survey (\cmo), constructed
   using the semi-analytic galaxy catalogues by De Lucia \& Blaizot
   (2007) applied to the Millennium Simulation. From
   each mock, we extracted a sub-sample of galaxies mimicking the VVDS
   observational strategy (\omo). We then computed the B-band
   Luminosity Function LF and the colour-density relation in mocks
   using the same methods employed for the VVDS data.}
   {We find that the B-band LF in mocks roughly agrees with the
   observed LF, but at $0.2<z<0.8$ the faint-end slope of the model LF
   is steeper than the observed one. Computing the LF for early and
   late type galaxies separately, we show that mocks have an excess of
   faint early type galaxies and of bright late type galaxies with
   respect to data. We find that the colour-density relation in \omos
   is in excellent agreement with the one in \cmo. This suggests that
   the VVDS observational strategy does not introduce any severe bias
   to the observed colour-density relation. At $z \sim 0.7$, the
   colour-density relation in mocks agrees qualitatively with
   observations, with red galaxies residing preferentially in high
   densities. However, the strength of the colour-density relation in
   mocks does not vary within $0.2<z<1.5$, while the observed relation
   flattens with increasing redshift and possibly inverts at
   $z\sim1.3$. We argue that the lack of evolution in the
   colour-density relation
 in mocks cannot be due only to inaccurate prescriptions for the
 evolution of satellite galaxies, but indicates that also the treatment of the
 central galaxies has to be revised.} 
   {The reversal of the colour-density relation can be explained by
   wet mergers between young galaxies, producing a starburst
   event. This should be seen on group scales, where mergers are
   frequent, with possibly some residual trend on larger scales. This
   residual is found in observations at $z=1.5$ on a scale of
   $\sim$5$h^{-1}$Mpc, but not in the mocks, suggesting that the
   treatment of physical processes influencing both satellites and
   central galaxies  in models should be revised. A detailed analysis
   would be desirable also on small scales, which requires flux limits
   fainter than those of the VVDS data.}  
   \maketitle


\section{Introduction}

Several physical mechanisms are expected to influence the 
properties of galaxies in over-dense regions: ram pressure stripping
of gas \citep{gunn_gott1972}, galaxy-galaxy merging
\citep{toomre1972}, strangulation \citep{larson1980}, and harassment
\citep{fk81,moore1996}. Each mechanism has specific environmental
dependencies and timescales, but their relative role in regulating
galaxy formation and determining the observed trends remains
unclear. In addition, it is not yet clear to what extent internal
processes such as feedback from supernovae and central black holes
contribute to the observed environmental dependence of the galaxy
structural parameters.  Finally, it is known that in a Gaussian random
field there is a statistical correlation between mass fluctuations on
different scales, with most massive halos preferentially residing
within large scale over-densities
\citep[see][]{kaiser1987,mo1996}. What is less clear is the role of
initial cosmological conditions in modulating the observed density
dependence of galaxy properties \citep{abbas2005}.

Semi-analytic models (SAMs) of galaxy formation coupled with dark
matter (DM) simulations provide a useful tool to address these issues.
In SAMs, the individual physical processes taking place during galaxy
evolution are expressed through simple equations that are motivated by
observational and/or theoretical studies. These equations parametrise
the dependency of such processes on physical properties of galaxies
and/or those of the dark matter haloes in which they reside. Then, the
comparison between galaxy properties in SAMs and in the observed data
gives important feedback on the validity of the prescriptions used in
the models and on the nature of the observed correlation between
galaxy properties and environment. Models are usually tuned to
reproduce some (sub)set of observational data in the local Universe,
most notably the observed local galaxy luminosity function (LF) and/or
mass function. Tracing back in time galaxy evolutionary paths, pushing
these studies at high redshifts, becomes thus a very powerful tool in
terms of discriminating between different models.

The recent completion of large and deep high-redshift surveys makes a
detailed comparison between model predictions and observational data
possible \citep[see
e.g.][]{Stringer_etal_2009,delatorre2011_MILLVVDS}.  Following galaxy
evolution over a large redshift range can help addressing the open issue of the apparent
contradiction between the hierarchical growth of DM structures and the
`downsizing' scenario of luminous matter. A number of
different observational tests support a `hierarchical' scenario for
structure formation in which DM halos form first in the highest
density peaks of the primordial density field, and then grow
hierarchically through subsequent mergers
\citep[e.g.][]{peebles1980_book}. Observations have shown that galaxy
evolution does not proceed in a similar `bottom-up' fashion, at least
for their star formation (SF) histories: with increasing cosmic time,
SF moves towards less massive galaxies
\citep[][]{cowie96,gavazzi96}. As discussed in
e.g. \cite{DeLucia_etal_2006}, these findings are not necessarily evidence of
`anti-hierarchical' growth of luminous matter as the `formation' of the
galaxy stellar population does not coincide with its assembly (see
\citealt{fontanot2009} for a more detailed discussion about
different manifestations of downsizing and comparison with
observational data).

Simulations can also help clarifying the relation between different
definitions that are commonly adopted for the `environment'. Indeed,
different quantities have been used in the literature to characterise
the local and/or global environment. The use of different
environmental definitions makes it very difficult to compare results
from different surveys, and at different cosmic epochs. Using
simulated galaxy catalogues allow us to have a common reference for
the environment, such as the underlying DM distribution.

In this paper, we use mock catalogues constructed from semi-analytic
models applied to the Millennium
Simulation\footnote{http://www.mpa-garching.mpg.de/galform/virgo/millennium/}
\citep{springel2005_MILL} to carry out a detailed comparison with the
observational results presented in \citet[][C06
hereafter]{cucciati2006}, based on the VIMOS-VLT Deep Survey
\citep[VVDS, see][]{lefevre2005a}. A detailed comparison between
observations and model data for galaxy number counts, redshift and
colour distribution, and galaxy clustering was presented in
\citet[][Paper I hereafter]{delatorre2011_MILLVVDS}. Here we focus on
the observed colour-density relation and its evolution.

The aims of our study are: (i) test the robustness of observational
results in C06 versus possible biases due to the observational
strategy adopted (e.g., does the VVDS sampling rate alter the strength
of the colour-density relation?); (ii) test galaxy evolution in mocks
(do we see the same environmental effects on galaxy properties in
mocks and observed data?); (iii) understand what `environment' is
being studied (e.g., which is the relation between the density traced
by galaxies and the underlying DM distribution?).

This paper is organised as follows. Section~2 gives a summary of the
VVDS data used in C06 and describes the mock galaxy catalogues used for the
model comparison. In Section~3, we compare the rest-frame B-band LF of
the VVDS data with that obtained using mock catalogues. Section~4
describes the local density and colour distributions in mocks, and
compares the colour-density relation in C06 with the one found in
mocks.  In Section~5 we take advantage of the available mocks to
analyse how the local density computed on a $5h^{-1}$Mpc scale
compares with the total halo mass in which galaxies reside. In
Section~6 we discuss our results, and we summarise them in Section~7.

Throughout this paper, we use the AB flux normalisation for both
observed data and mocks. When we refer to observed data, we adopt the
concordance cosmology ($\Omega_{m}$, $\Omega_{\Lambda}$, $h$)~=~(0.3,
0.7, 0.7).


\section{Data and mock catalogues}\label{data_mocks}

\subsection{The VVDS Deep sample}\label{VVDSdata}

The VVDS is a large spectroscopic survey, with the primary aim of
studying galaxy evolution and large scale structure formation over the
redshift range $0<z<5$.  Details on the survey strategy are described
in \cite{lefevre2005a}.  VVDS is complemented by ancillary deep
photometric data: {\it BVRI} from the CFHT-12K camera
\citep{mccracken2003,lefevre2004b}, {\it JK} from the NTT telescope
\citep{iovino2005,temporin2008}, $U$ from the MPI telescope
\citep{radovich2004}, {\it $u^{*}$, g', r', i', z'}-band data from the
CFHT Legacy Survey and {\it JHK$_S$} from the WIRDS survey
\citep{bielby2011_WIRDS} with the CFHT-WIRCAM camera (see
\citealp{cucciati2012_sfrd} for a detailed description).

This paper is based on the colour-density relation studied in C06,
over the VVDS-0226-04 Deep field (from now on ``VVDS-02h field''). We 
refer the reader to that paper for a detailed description of the data.
Briefly, the VVDS-02h data set is a
purely flux limited spectroscopic sample, with $17.5 \leq I_{AB} \leq
24.0$. In this range of magnitudes, the parent photometric catalogue
is complete and free from surface brightness selection effects
\citep{mccracken2003}. Spectroscopic observations were carried out at
the ESO-VLT with the VIsible Multi-Object Spectrograph (VIMOS) using
the LRRed grism. The {\it rms} accuracy of the redshift measurements
is $\sim 275$ km/s \citep{lefevre2005a}. The VVDS-02h field covers a
total area of $0.7 \times 0.7$ deg$^2$, targeted by 1, 2 or 4
spectrograph passes, and it probes a comoving volume (up to \z=1.5) of
nearly $1.5 \times 10^6$ $h^{-3}$Mpc$^{3}$ in a standard $\Lambda$CDM
cosmology. The covered field has transversal dimensions $\sim$
37$\times $37 \hpc at \z=1.5.  Averaging over the area observed,
spectra have been obtained for a total of 22.8\% of the photometric
sources, and $\sim80$\% of these targeted objects yield a reliable
redshift, resulting in an overall sampling rate of $\sim18$\%
\citep[see][]{ilbert2005}.  The final galaxy sample used in C06
contains 6582 galaxies with reliable redshifts.

\subsection{The mock catalogues}\label{mocks_sec}

Mock galaxy catalogues were obtained by applying the semi-analytical
model of galaxy evolution described in \cite{delucia_blaizot2007} to
the dark matter halo merging trees derived from the Millennium
Simulation \citep{springel2005_MILL}. This contains $N = 2160^3$
particles of mass $8.6 \times 10^8 h^{-1} M_ {\odot}$ within a
comoving box of size 500 $h^{-1}$Mpc on a side. The adopted
cosmological model is a $\Lambda CDM$ model with $\Omega_m = 0.25$,
$\Omega_b = 0.045$, $h = 0.73$, $\Omega_{\Lambda}= 0.75$, $ n = 1$ and
$\sigma_8 = 0.9$.

The semi-analytical model used here is fully described in
\cite{delucia_blaizot2007}. It builds on results from previous works
\citep{kauffmann00_SAM,springel01_SAM, delucia04_SAM,croton2006} to
describe the relevant physical processes, e.g. star formation, gas
accretion and cooling, formation of super-massive black holes and
galaxy mergers, feedback (also `radio mode' feedback). The SAM model
used in this study has been tested against many observational data and
and it has been shown to provide a relatively good agreement with
observational data both for the local Universe and at higher redshift.
It is, however, not without problems.  In particular, the fraction of
low-mass red galaxies is too high, and the clustering signal of red
galaxies is overpredicted. The galaxy mass function has a too high
normalisation in the regime of low and intermediate stellar masses, at
any redshifts. We refer the reader to \cite{weinman2006b},
\cite{wang2008}, \cite{fontanot2009} and Paper I for more details. It
is worth noticing that these problems are not specific of the
particular model used here but appear to be common to most recently
published SAMs. We will further discuss some of these discrepancies in
the following sections.

For our analysis, it is important to summarise how absolute magnitudes
are computed in the SAM adopted here. The \cite{BC03} model is used to
generate lookup tables for the luminosity of a single burst of fixed
mass, as a function of the age of the stellar population ($t$) and its
stellar metallicity ($Z$). At each star formation episode, the model
interpolates between these tables, using a linear interpolation in $t$
and log$(Z)$, to calculate the contribution to the luminosity of model
galaxies at the time of the observations. Stars are assumed to form
with the metallicity of the cold gas component, and an instantaneous
recycling approximation is adopted. The model adopts a
\cite{chabrier2003_IMF} IMF with a lower and upper mass cut-off of
0.1 and 100 $M_{\odot}$, respectively. Magnitudes are extincted (internal
extinction) using the model that is detailed in \cite{delucia_blaizot2007}.

For our study, we randomly selected 20 $1\times 1$ deg$^2$ Millennium light cones, from those
constructed using the code MoMaF \citep{Blaizot2005} as described in Paper
I. From these cones, we extracted two sets of mock catalogues. First, we
extracted $1\times 1$ deg$^2$ mocks with the same flux limits as the
VVDS-02h sample ($17.5 \leq I_{AB} \leq 24$), and added {\it a
posteriori} the same redshift measurement error of the VVDS
sample. These catalogues, having 100\% sampling rate, will be called
from now on \cmo. Then, from the \cmos we extracted mock catalogues
mimicking our VVDS-02h sample, i.e. adding geometrical effects and
uneven sampling rate.  We call these mocks \omo. We refer the reader
to Paper I for details.

As we will show later (Sec.  \ref{mock_counts_sec}), Millennium light
cones have on average higher $I$-band number counts (per surface area
and over the range $17.5 \leq I_{AB} \leq 24$) than VVDS. In this
study we complement the 20 random light cones with two additional
cones with the lowest number counts (cones 036d and 045b), among those
presented in Paper I. VVDS number counts are in between the number
counts of these two cones. We also added the cone with the highest 
number counts (cone 022b).

\begin{figure} \centering
  \includegraphics[width=9cm]{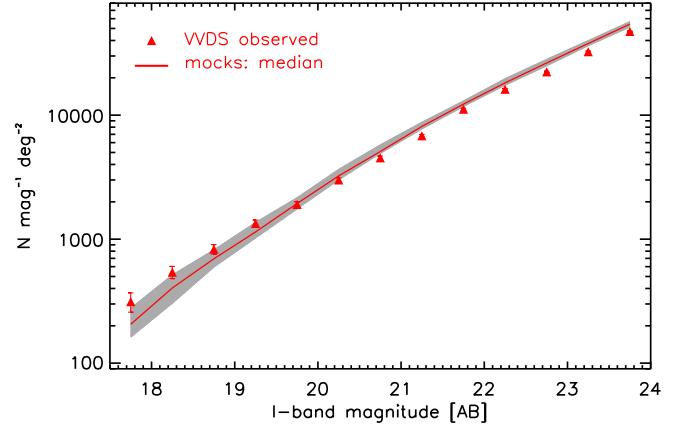}
  \caption{Galaxy number counts per unit magnitude and per square
    degree. Red triangles are for VVDS measurements, with error bars
    representing the Poissonian uncertainty. The red line shows the
    median number counts of the 23 mocks, and the grey area shows the
    16$^{th}$ to 84$^{th}$ percentile range of the mock number counts
    distribution. The VVDS number counts are in good agreement with
    number counts obtained by \cite{mccracken2003} for the
    CFH12K-VIRMOS deep field.}
\label{counts} \end{figure}

\begin{figure} \centering
\includegraphics[width=9cm]{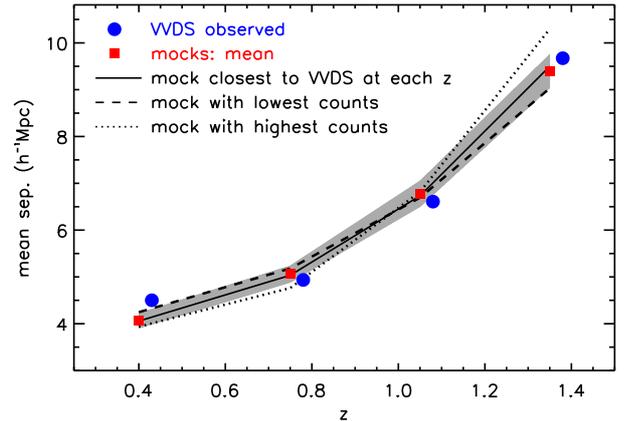}
\caption{Mean inter-galaxy separation as a function of redshift, in
  the four redshift bins used in C06 (0.25-0.6, 0.6-0.9, 0.9-1.2, 1.2-1.5). Red squares
  are the mean among the 23 \omo~used in this study, and the grey
  shaded area shows the $1-\sigma$ scatter.  Blue circles are for VVDS
  data. The thick solid line corresponds to the mock with mean
  inter-galaxy separation closest to the VVDS measurements at all
  redshift bins considered. The dashed line corresponds to the mock
  with lowest number counts (in the total redshift range 0.25-1.5),
  and the dotted line to the cone with highest number counts.}
\label{mean_sep}
\end{figure}

\subsubsection{Mock catalogues number counts and mean inter-galaxy
  separation}\label{mock_counts_sec}

Since we will compute the galaxy 3D local environment in mocks, and
compare it to the one found in the VVDS, it is important to know
possible similarities and/or discrepancies between the 3D galaxy
distribution in the mocks and that in the data.

In Paper I, we have shown that the light-cones used in our study give
an average redshift distribution $n(z)$ and galaxy clustering that are
not in perfect agreement with the VVDS-02h measurements. In
particular, we showed that the mock $n(z)$ agrees with the observed
one in the range $0.5<z<1.8$, but it overestimates the observed number
of galaxies at $z<0.5$, and underestimates it at $z>1.8$. In addition,
we have shown that the Munich semi-analytical model overestimates the
VVDS clustering, and that this is mainly due to an excess of the
clustering signal for model red galaxies.

Fig.~\ref{counts} shows the number counts per square degree of
VVDS-02h galaxies as a function of the observed $I_{AB}$ magnitude
(that is the VVDS selection magnitude). Number counts have been
computed from the parent photometric catalogue of $\sim$40,000
sources, with stars removed using the method described in
\cite{mccracken2003}. The median number counts of the 23 mocks used in
this study are over-plotted as a solid line. Fig.~\ref{counts} shows
that the mock $I$-band number counts are higher than observational
data at $I_{AB}>22$, but lower than the observed number counts at the
brightest magnitudes ($I_{AB}<$20). Since faint galaxies are much more
numerous than bright ones, mocks contain on average $\sim 10$\% more
galaxies than observations\footnote{Fig.~3 of Paper I shows that
  $I$-band counts in the mocks are in very good agreement with those
  from the VVDS. The discrepancy with results presented here is
    due to an error in the plotting routine used for that Figure.
  This however does not alter the conclusions of that section of Paper
  I, which are based on the $i'$-band counts (Fig. 4 in paper I): the
  $i'$-band counts in the mocks are higher than in the VVDS,
  consistently with our $I$-band results.}.

Fig.~\ref{mean_sep} shows the mean inter-galaxy separation in the
VVDS-02h field, as a function of redshift. We
overplot the same quantity computed in the \omo. In the lowest
redshift bin, all mocks considered have mean separation lower than
measured in observed data. The average mean separation in mocks is
larger than the VVDS one in the range $0.6<z<1.2$, and lower than the
measured VVDS value for $z>1.2$. At $z>0.6$, there is at least one
mock with mean separation close to the VVDS one, but this is not the
same mock at all redshifts. In Fig.~\ref{mean_sep}, we also show the
mean inter-galaxy separation for the mock with the lowest number
counts (dashed line). Its mean inter-galaxy separation is larger or
very similar to the VVDS one (at least up to $z=1.2$), as one would
expect. The opposite is true for the mock with the highest number
counts (dotted line), up to $z=1.2$.  Among those used in this study,
there is no mock with number counts close to the observed ones at all
redshift bins. The black solid line in the figure shows the mock that
has inter-galaxy separation on average the most similar to VVDS at all
redshift bins. Its counts are very close to the average among all the
mocks at each redshift (red squares). It does not correspond to the
mock with the lowest total number counts.

The results on the $n(z)$, the number counts as a function of
selection magnitude, and the mean inter-galaxy separation are all
related, showing that \omos contain more galaxies than the VVDS-02h
field.

\subsubsection{The absolute magnitudes}\label{absmag_comp}

In C06, red and blue galaxies are defined using the rest-frame colour
$u^{*}-g'$. The $u^{*}$ and $g'$ (CFHT-LS filters) rest-frame
magnitudes are not available on the Millennium database. Among the
rest-frame absolute magnitude available for the
\cite{delucia_blaizot2007}, the closest to $u^{*}$ and $g'$ are the
ones in the B and V Johnson filters. To be consistent with the
analysis discussed in C06, we computed for our mocks the $u^{*}$ and
$g'$ absolute magnitudes using the same method employed for the VVDS
data (see below). We also computed for our mocks the absolute
magnitudes in the B and V Johnson filters, so as to compare them with
the intrinsic ones available from the simulations. In this way, we can
verify that the rest-frame magnitudes computation does not affect the
intrinsic luminosity distribution.

For the VVDS data set, we computed the absolute magnitudes using the
code ALF (Algorithm for Luminosity Function, \citealp{ilbert2005}),
based on a SED fitting technique. To reduce the dependency on
templates, we derive the rest frame absolute magnitude in each band
using the apparent magnitude from the closest observed band, shifted
at the redshift of the given galaxy.  In this way, we minimise the
applied K-correction, that depends on the assumed template. The set of
observed magnitudes used to derive the absolute magnitudes in C06
included {\it BVRI} bands from the CFHT-12K camera, and {\it $u^{*}$,
g', r', i', z'}-band data from the CFHT Legacy Survey. We have these
observed magnitudes in our mocks, and we
used them to compute the intrinsic luminosities in the mocks, with the
code ALF. We adopted the same cosmology used for the VVDS data,
i.e. $\Omega_m = 0.3$, $\Omega_{\Lambda}= 0.7$, $h = 0.7$.

We compared the B and V absolute magnitudes (not including internal
dust extinction) available on the Millennium database with the corresponding
quantities computed with ALF.  We found that the two sets of
magnitudes are consistent within $\sim5\%$. The small difference
between the two sets of magnitudes is slightly dependent on the galaxy
luminosity, and does not depend on redshift within the redshift range
explored. As B and V filters are very close to $u^{*}$ and $g'$, we
are confident that also the computation of $u^{*}$ and $g'$ rest-frame
magnitudes using the method described above does not introduce any
spurious effect. 

\begin{figure*} \centering
\includegraphics[width=0.48\hsize]{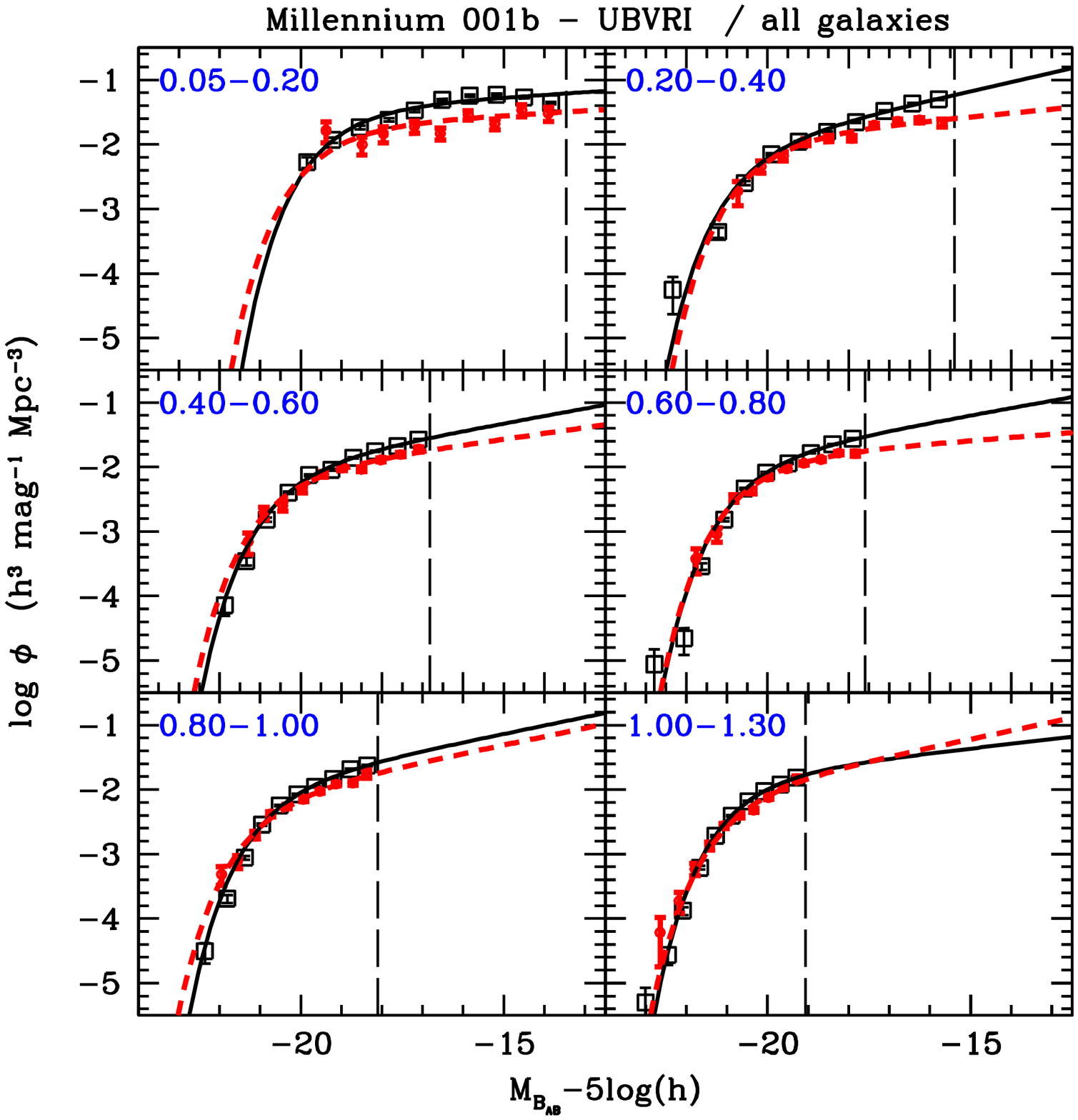}
\includegraphics[width=0.48\hsize]{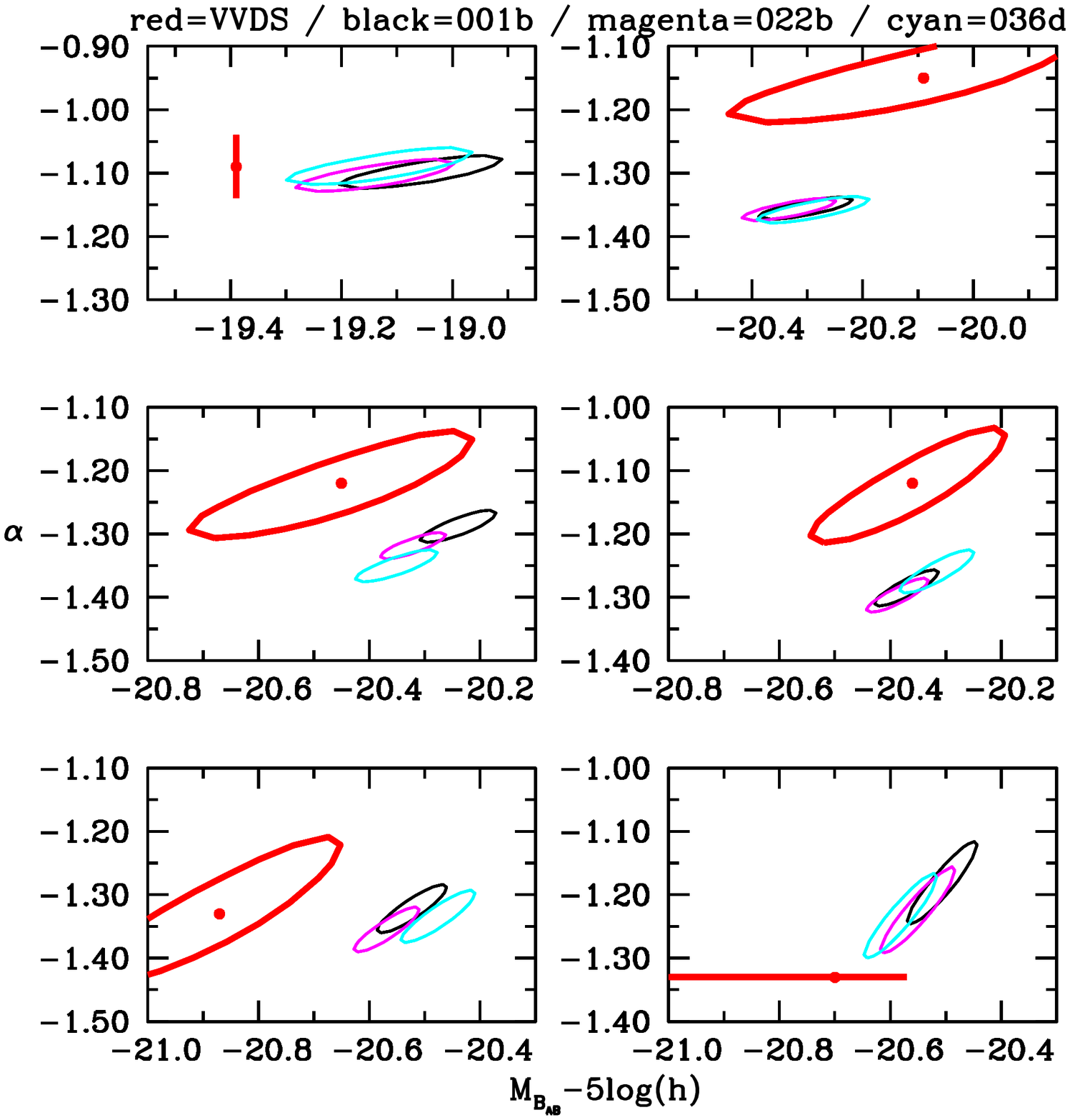}
\includegraphics[width=0.48\hsize]{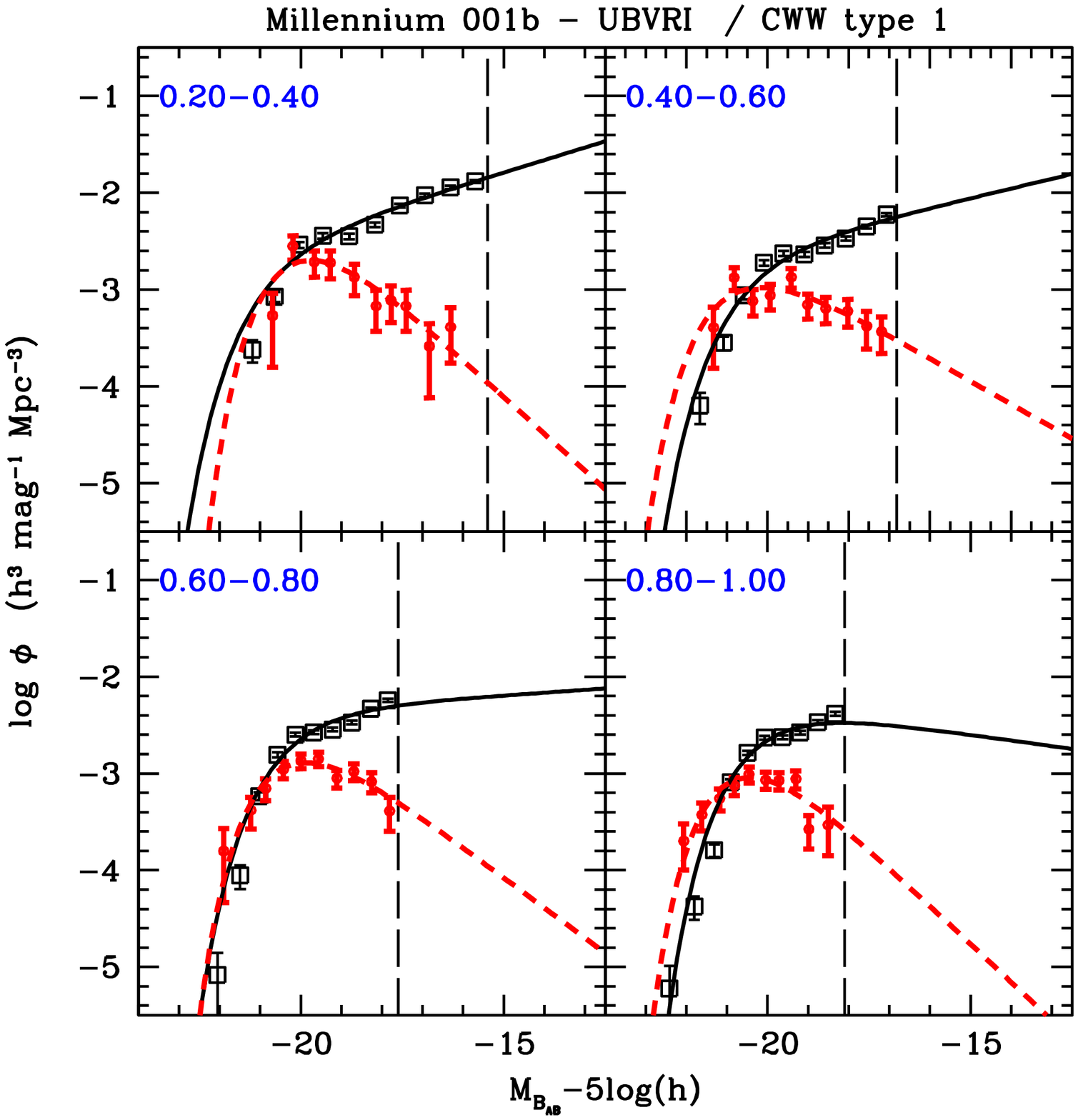}
\includegraphics[width=0.48\hsize]{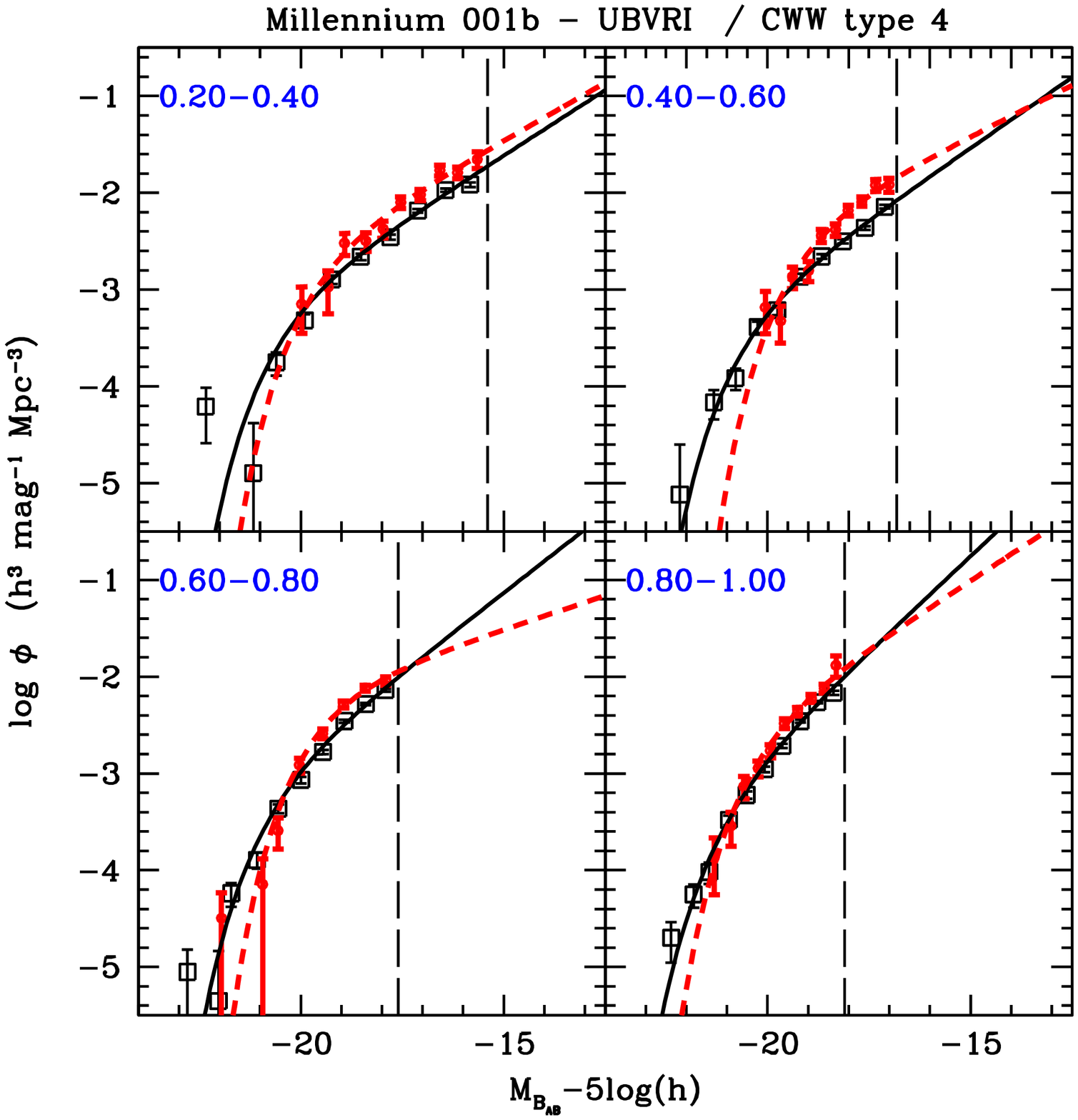} 
\caption{Evolution of the luminosity function of galaxies in the mock samples,
  compared to the results from the VVDS.  {\it Upper left panel:} Evolution of
  the luminosity function in the B-band for all galaxies in the \ucmos 001b.
  Each panel refers to a different redshift bin, which is indicated in the
  label. The vertical dashed line represents the faint absolute limit
  considered in the $STY$ estimate. The luminosity functions are estimated with
  different methods (see text for details) but for clarity we plot only the
  results from $C^+$ (symbols) and $STY$ (lines). Red filled circles and lines are
  used for observational measurements while black empty squares and lines are used
  for the corresponding measurements from the mock. {\it Upper right panel:}
  $68\%$ confidence ellipses for the $\alpha$ and $M^*$ parameters obtained
  using all galaxies in mock 001b (black ellipse), mock 022b (magenta ellipse),
  and mock 036d (cyan ellipse). The red thicker ellipse corresponds to the VVDS
  measurements; the red thick lines in the first and the last redshift bins show the
  uncertainties on one of the parameters obtained keeping the other fixed.
  {\it Lower left panel:} Evolution of the luminosity function in the B-band
  for type 1 galaxies in the mock 001b. Lines and symbols have the same
  meaning as in the upper left panel.  {\it Lower right panel:} Evolution of
  the luminosity function in the B-band for type 4 galaxies in the mock
  001b. Lines and of symbols have the same meaning as in the upper left panel.}
\label{lumfun} 
\end{figure*}


\section{The B-band Luminosity Function}\label{LFsec}

In this section, we compare the VVDS B-band (Johnson) Luminosity
Function presented in \cite{ilbert2005} and \cite{zucca2006_VVDS_LF}
with the corresponding quantity derived from the mocks. In
\cite{zucca2006_VVDS_LF}, we used only $UBVRI$\footnote{$U$-band from
the MPI telescope, and $BVRI$ from the CFHT-12K camera as described in
Sec.~\ref{VVDSdata}} apparent magnitudes as input for the SED fitting,
because the CFHT-LS photometry was then not yet available. For
consistency, we computed a second set of absolute magnitudes for the
mocks, using only these five observed bands. This second set of
absolute magnitudes has been used only for the LF presented in this
section. For this analysis, we have used the \cmos catalogues, and
compared model results with the observational estimate corrected for
incompleteness.

The LF was computed using the code ALF (see Sect.~\ref{absmag_comp}),
which implements several estimators: the non-parametric $1/V_{max}$
\citep{schmidt68_Vmax}, $C^+$ \citep{Lynden_Bell_LF_Cplus}, $SWML$
\citep{efstathiou88_LF_SWML}, and the parametric $STY$
\citep{sandage79_LF_STY}. We used the $STY$ assuming a single Schechter
function \citep{schechter1976} parametrised in terms of a
characteristic luminosity ($L^{*}$), a faint-end slope ($\alpha$), and
a normalisation (density) parameter ($\phi^{*}$). For a more detailed
description of the tool and the estimators, we refer to
\cite{ilbert2005}.

\cite{ilbert2004} have shown that the LF measurement can be biased,
mainly at the faint end, when the band used is far from the rest frame
band in which galaxies are selected. This is due to the fact that,
because of the K-correction, galaxies of different type are visible in
different absolute magnitude ranges at a given redshift, even when
applying the same flux limits. Moreover, in a flux-limited survey,
this limit varies with redshift. When computing the VVDS LF, we
avoided this bias by using in each redshift range only galaxies within
the absolute magnitude range where all the SEDs are observable. We
have computed the LF in mocks using the same method.

The upper left panel of Fig.~\ref{lumfun} shows the luminosity
function for one of the \cmos catalogues (cone 001b), in different
redshift bins, obtained with $C^+$ and $STY$ methods.  The luminosity
functions derived with the other two methods ($1/V_{max}$ and $SWML$)
are consistent with those shown in the figure.  The dashed red line
and red filled circles in each panel show the corresponding results
for the VVDS data. The vertical dashed line represents the faint
absolute limit considered in the $STY$ estimate.

The upper right panel of Fig.~\ref{lumfun} shows the confidence
ellipses of the $\alpha$ and $M^*$ parameters obtained in \ucmos 001b
(black ellipse), and also in \ucmos 022b (magenta ellipse) and \ucmos
036d (cyan ellipse), which are the mocks with the lowest and highest
number counts (see Sect.~\ref{mocks_sec}). The red thicker ellipse
refers to the VVDS parameters, and the red thick lines in the first
and last redshift bins show the uncertainties on one of the parameters
derived by keeping the other one fixed. This figure shows that shape
parameters $\alpha$ and $M^*$ measured for the three mocks considered
are consistent within the uncertainties. Some differences can be found
in the normalisation parameters $\phi^*$, reflecting the different
number counts in the mocks.

Overall, there is a reasonable agreement between the LF in the VVDS
and in mocks, but with some non negligible discrepancies. In
particular, we find that $M^*$ is almost always brighter and $\alpha$
almost always flatter in the VVDS than in the mocks. In addition,
there are significant differences in the normalisation of the model
and data LFs, reflecting the differences in the number counts
discussed in Sect.~\ref{mock_counts_sec}.  In order to better explore
these differences and to understand if they are induced by a specific
class of objects, we derived the LF for galaxies of different types.

\cite{zucca2006_VVDS_LF} split the global galaxy population in
different {\it spectro-photometric types}. For each galaxy, they found
the best template fitting the galaxy SED, choosing among four
empirical templates from \citet[][CWW hereafter]{CWW1980} and two
starburst templates computed using GISSEL \citep{BC1993}. They defined
four galaxy types, corresponding to the four CWW templates (E/S0,
early spiral, late spiral and irregular - type 1, 2, 3, and 4
respectively). Type 4 galaxies include the two starburst templates.

Here we apply the same classification scheme to galaxies in the mocks,
and derive their LFs as done in \cite{zucca2006_VVDS_LF}. The lower
panels of Fig.~\ref{lumfun} shows the LFs obtained for the two extreme
galaxy populations (type 1 in the left bottom panel, and type 4 in the
right bottom panel) in the mock 001b and in the VVDS data. Results
from the mocks with largest and lowest number counts are similar. The
figure shows a clear excess of type 1 galaxies in the mocks at
magnitudes fainter than the `knee' of the LF. In contrast, bright type 1
galaxies are under-represented in mocks. For the
latest-type galaxies (type 4), in mocks there is an excess of bright
galaxies in the lowest redshift bins considered, and a slight deficit
of fainter galaxies.  Therefore, the steeper faint-end slope and
the fainter $M^{*}$ found in the global LF in mocks are due to an
excess of faint type 1 galaxies and deficit of
bright type 1 galaxies, respectively. The effect on the global LF of
the deficit of bright type 1 galaxies is less evident because
it is compensated by the excess of bright type 4 galaxies.

The excess of faint red galaxies in the model used in this paper is a
known problem which actually appears to be shared by most (all)
semi-analytic models that have been published in recent years
\citep[see e.g.][and references
therein]{wang2007_model,fontanot2009,Weinmann_etal_2010}. This
is also consistent with what we found in Paper I.


\section{The colour-density relation}

In C06, we analysed the colour-density relation as a function of both
redshift and galaxy luminosity. Namely, we split our sample in four
redshift bins ($0.25 \leq \z < 0.6, 0.6 \leq \z < 0.9, 0.9 \leq \z <
1.2$ and $1.2 \leq \z < 1.5$), and in each redshift bin we studied the
colour-density relation for galaxies with $(M_B-5 \log h) \leq -19.0,
-19.5, -20.0, -20.5, -21.0$.

We stress that the density has been computed using the entire sample
available, irrespectively of galaxy luminosity (see
Sect.~\ref{environment}). Moreover, at $0.25< \z \leq 0.6$, the VVDS
sample does not contain enough galaxies with $(M_B-5 \log h) \leq
-20.5$ because of the small probed volume. So we excluded the two
brightest luminosity thresholds from the analysis in this redshift
range.  We also know that VVDS samples brighter and brighter galaxies
at higher redshift, due to its flux limit. So we examined only
galaxies with $(M_B-5 \log h) \leq -19.5$ and $\leq -20.0$ at $0.9
\leq \z \leq 1.2$ and $1.2 \leq \z \leq 1.5$, respectively. These
lower luminosity limits assure that the considered sub-samples are
complete for all galaxy types.

\subsection{The environment parameterisation}\label{environment}

We refer the reader to C06 for a detailed description
of the density computation method. Here we only give a brief
summary. 

For each galaxy at a comoving position {\bf r}, C06
characterised the environment that surrounds it by means of the
dimensionless 3D density contrast $\del({\bf r},R)$, smoothed with a
Gaussian filter of dimension $R$: $\del({\bf r},R) = [\rho({\bf
r},R)-\overline{\rho}({\bf r})] / \overline{\rho}({\bf r})$. When
smoothing, galaxies are weighted to correct for various survey
observational characteristics (sample selection function, target
sampling rate, spectroscopic success rate, and angular sampling
rate). Moreover, underestimates of $\delta$ due to the presence of
edges have been corrected by dividing the measured densities by the
fraction of the volume of the filter contained within the survey
borders.

In C06, we calibrated the density reconstruction scheme using
  simulated mock catalogues extracted from GalICS \citep{Hatton2003}.
  The aim was to determine the redshift ranges and smoothing length
  scales R over which our environmental estimator reliably reproduced
  the underlying galaxy environment, as given by a 100\% sampling rate
  catalogue with $I_{AB}\leq 24$. We concluded that we reliably
  reproduced the underlying galaxy environment on scales $R\ge 5$\hpc
  out to z=1.5.  In the present study, we do not use GalICS mocks
  because they do not provide the information needed for our analysis
  (e.g. the observed magnitudes in all the VVDS bands). However, for a
  sanity check, we repeated the tests carried out in C06 using the
  mocks used in the present study, and we confirmed our previous
  results on the reliability of the density reconstruction.

We computed the density field in both the \cmos and \omo, using the
same method as in the VVDS data, and the same flux-limited
($I_{AB}\leq 24$) tracers population. As mentioned in
Sec.~\ref{mocks_sec}, for the \omos we have adopted the same weighting scheme used
for the VVDS data. In particular, we have considered the
target sampling rate, the spectroscopic success rate, and angular
sampling rate, and also accounted for survey boundaries. By
construction, no correction is needed for the \cmos since they have a
$100$\% sampling rate. Moreover, to compute the density in the \cmo,
we started from the 1x1 deg$^2$ mocks from which they were extracted,
so the density field depends very little (if any) on the correction
for boundary effects.

\begin{figure}
\centering
\includegraphics[width=9cm]{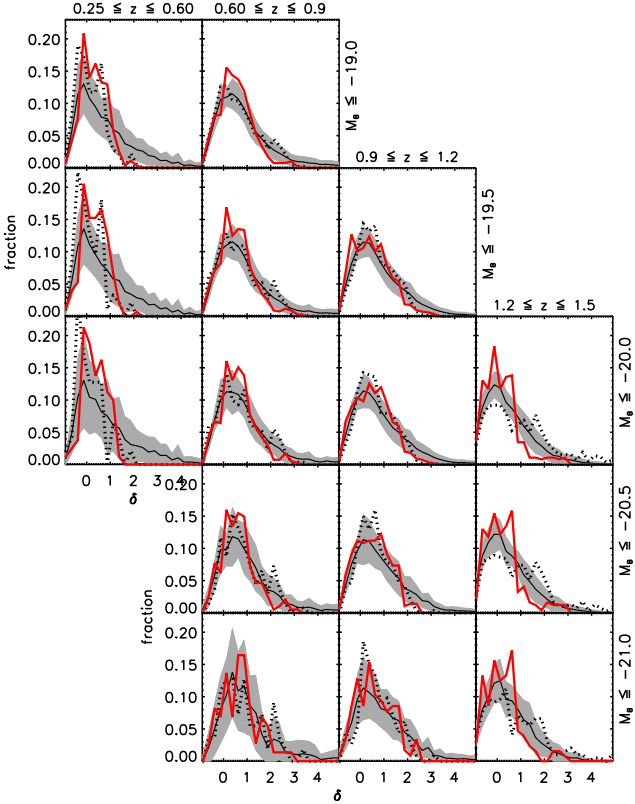}
\caption{Density contrast distribution in the different redshift bins
  and luminosity limits explored. Red thick solid line: VVDS sample;
  black dotted line: \uomos with lowest number counts; black thin solid line: average of the
  \omo; grey area:
  1-$\sigma$ scatter of \omo.  }
\label{delta_distrib}
\end{figure}

Fig.~\ref{delta_distrib} shows the density contrast distribution for
\omos and VVDS data, for the different redshift bins and luminosity
limits explored in C06. The average density distribution in \cmos is
very close to that in \omo, but \cmos have a smaller scatter around
the mean.  The density distribution in mocks has longer tails towards
high densities than VVDS, in particular at the lowest and highest
redshift explored. Interestingly, these longer tails are not
consequence of the higher number counts in mocks: the figure shows
that also the mock with lowest number counts exhibit these tails at
high density. Moreover, we verified that, depopulating randomly the
\omos to have them matching the observed number counts as a function
of $I$-band, and then re-computing $\delta$, does not suppress these
tails. This means that in the Munich semi-analytical model, the 3D
spatial distribution of galaxies is intrinsically different from that
observed in the real Universe, and that the excess of the clustering
signal in the mocks is only in part due to an excess of low- to
intermediate-mass galaxies.  In Paper I, we
  suggested that this might be due to the assumption of a WMAP1
  cosmology in the Millennium Simulation, and in particular to the use
  of a high normalisation of the power spectrum ($\sigma_8$). A lower
  value of $\sigma_8$ would reduce the overall density contrast at any
  given redshift \citep{wang2008}. Indeed, in Paper I we showed that,
  converting the correlation functions in the model to those expected
  assuming a lower value of $\sigma_8$, the signal decreases at all
  scales. However, in Paper I we did not change  other cosmological
  parameters. Recently, \cite{guo2012_millWMAP7} rescaled the
  Millennium simulation to the WMAP7 \citep{komatsu2011_WMAP7}
  cosmological parameters values. They showed that the effects of the
  decreased $\sigma_8$ are compensated by the higher value of
  $\Omega_{m}$. Therefore, the assumption of a WMAP1 cosmology cannot
  explain the excess of clustering in our mocks.

\begin{figure}
\centering
\includegraphics[width=9cm]{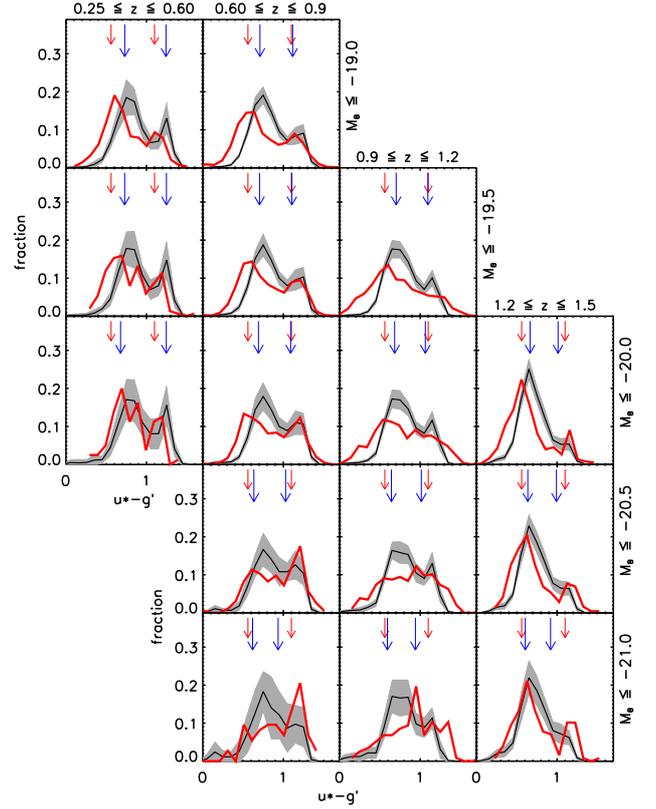}
\caption{Colour distribution in the different redshift bins and
  luminosity limits explored. Red thick solid line: VVDS sample;
   black thin solid line: average of the \omo; grey area: 1-$\sigma$
   scatter of \omo. Red short 
  arrows represents the fixed colour cuts used to define red 
  ($(u^{*}-g') \geq 1.1$) and blue ($(u^{*}-g') \leq 0.55$)
  galaxies in C06. Blue long arrows indicates which colour
  cuts should be used in the mocks to have the same percentage of red
  and blue galaxies (irrespectively of environment) as in the VVDS,
  for each redshift bin and luminosity limit. }
\label{col_distrib}
\end{figure}

\begin{figure} \centering
\includegraphics[width=9cm]{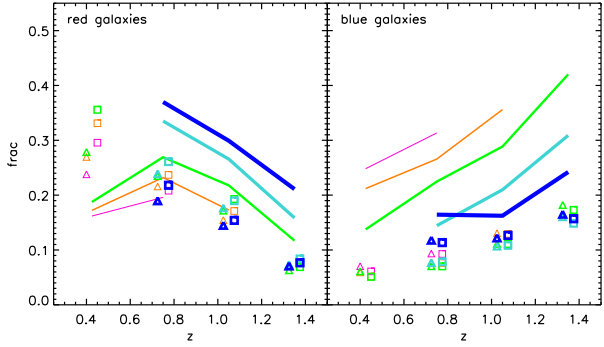}
\caption{Fraction of red ($(u^{*}-g') \geq 1.1$, left
    panel) and blue ($(u^{*}-g') \leq 0.55$, right panel) galaxies as
    a function of redshift, for different luminosity threshold. Lines
    are for VVDS data, triangles for \omos and squares for
    \cmo. The colour/thickness code is the following: from
    magenta/thin to blue/thick we consider galaxies with $(M_B-5 \log
    h) \leq -19.0, -19.5, -20.0, -20.5, -21.0$. Each triangle(/square)
    represents the mean value of all the \omo(/\cmo).}
\label{col_frac} 
\end{figure}

\subsection{The colour distribution}\label{color_distr_sec}

C06  empirically defined red and blue galaxies to be the two
extremes of the {\it $u^{*}$-g'} colour distribution. Namely, red
galaxies are defined as those with $u^{*}-g'\geq 1.1$ and blue as those with
$u^{*}-g'\leq 0.55$. These colour cuts roughly correspond to the two
peaks of the bimodal colour distribution in the VVDS. C06 kept these
limits fixed for all luminosities and redshift bins considered.

From Paper I and from Sect.~\ref{LFsec}, we know that the colour
distribution in mocks is different from the observed one.
Fig.~\ref{col_distrib} shows the colour distributions in \omos and
VVDS for the different redshift bins and luminosity limits considered
in this paper. The 1-$\sigma$ scatter for the \cmos is narrower than
the \omos one, but the average values of the two kinds of mocks are
very close. We note that at intermediate redshifts ($0.6<z<1.2$), the
blue cloud is more populated in the mocks than in the VVDS, especially
for bright galaxies ($M_B \leq -20$). At these redshifts and
luminosities, the blue peak of the bimodal colour distribution in the
mocks is from 30\% to 100\% higher then the VVDS one, while having
roughly the same width. Still, if we consider only the tail of bluest
galaxies ($u^{*}-g'\leq 0.55$, as defined in C06), these are less in
the mocks than in the VVDS, especially for faint galaxies. In
contrast, the peak of the red galaxy population is higher in the mocks
than in data for fainter galaxies, and it corresponds to redder
colours. These results are in agreement with those presented in Paper
I, and with the discrepancies discussed in Sect.~\ref{LFsec}.

Therefore the mix of galaxy populations (and colours) is different in
mocks and VVDS. As a consequence, if we used the same colour cuts in
the mocks as in the VVDS, we will have different fractions of red and
blue galaxies (independently of environment).  Fig.~\ref{col_frac}
shows the fraction of red and blue galaxies as defined in C06 in the VVDS, as a
function of redshift and for the different luminosity limits
considered. The corresponding fractions in the mocks are different from those in
the VVDS sample. Also,
the colour dependency on luminosity is less evident
in the mocks than in the VVDS. 

Fig.~\ref{col_frac} also shows that the fraction of red galaxies is
slightly lower in \omos than in \cmos at all redshifts, especially at
$z<0.6$, where it is lower by 5-8\% depending on the luminosity cut.
It is known that the VVDS observational strategy misses a very small
fraction ($\sim4$\%) of very red galaxies in particular at low $z$ (see e.g.
\citealp{franzetti2007}). This bias is mainly due to the optimisation
for slit positioning, that targets preferentially smaller galaxies in
$I$-band. At low redshift, brighter and bigger galaxies in the observed
$I$-band are the early types.  
We will show that this small loss of red galaxies at $z<0.6$ does not
alter the colour-density relation in \omo, and we believe that this
applies to the observed data as well. We also verified that the global
sampling rate does not vary as a function of the density computed on a
$5h^{-1}$Mpc scale in the \omo, for the redshift bins and luminosity
cuts considered.

Fig.~\ref{col_distrib} also shows the two colour cuts of C06 (red
arrows), and the colour cuts we should use in the mocks in order to
have the same fractions of red and blue galaxies as in the VVDS for
each redshift bin and luminosity cut (blue arrows). For
this plot we are considering all galaxies, independently of their
environment. To compute the colour-density relation in mocks, we have
taken the colour cuts corresponding to the same fraction of red and
blue galaxies as in the VVDS, as this choice follows the same
rationale used in C06, \ie considering the extremes of the colour
distribution. This makes the comparison with VVDS more
straightforward, and provides the same overall normalisation in the
colour-density relation as in the VVDS. As a test, we have also
computed the colour-density relation in the mocks using colour cuts
that correspond roughly to the two peaks of the mock colour
distribution. As for the VVDS, we kept these cuts constant at all
redshifts and for all luminosities. The colour-density relations
obtained in this way are very similar to those obtained using variable
cuts that reproduce the observed fraction of red and blue galaxies. So
our results are robust against this choice.

\subsection{The colour-density relation up to $z\sim 1.5$ }

We reproduced the analysis on the colour-density relation as in C06, using both
\cmos and \omos catalogues.

\begin{figure} \centering
\includegraphics[width=\hsize]{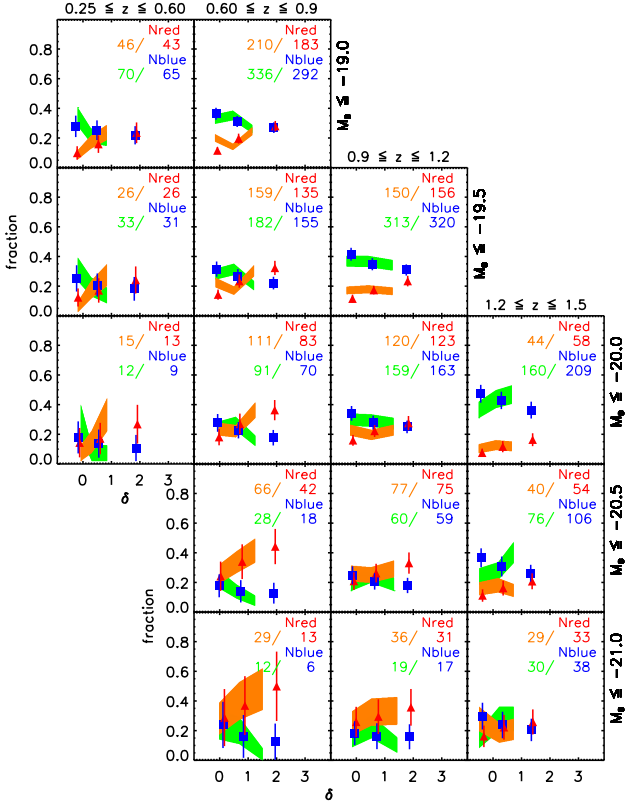}
\caption{Colour-density relation in \omo: blue squares represent the
  fraction of blue galaxies ($f_{blue}$) and red triangles the fraction of red
  ones ($f_{red}$). {\it x-} and {\it y-}axis positions are mean values of the
  {\it x-} and {\it y-}axis values for the 23 mocks. Vertical error
  bars includes the scatter among the 23 cones and the typical (mean)
  error in each mock on the computation of $f_{blue}$ or $f_{red}$.
  Orange and green contours represent $f_{red}$ and $f_{blue}$ 
  in the VVDS sample (from C06). The numbers quoted in each panel
  are the total number of red and blue galaxies in the VVDS data
  and in the mocks (average on the 23 mocks), with the same colour code
  as for the symbols. For VVDS observed data we define red
  galaxies those with $(u^{*}-g') \geq 1.1$ and blue those with
  $(u^{*}-g') \leq 0.55$, at all $z$ and for all luminosities. In
  the mocks, the definition of red and blue galaxies varies with $z$ and
  luminosity, in order to have the same fraction of red and blue
  galaxies as in the VVDS (irrespectively of environment) in each
  redshift bin and for each luminosity limit (see the blue arrows in
  Fig.~\ref{col_distrib}).}  
\label{col_den_2h} 
\end{figure}

\begin{figure} \centering
\includegraphics[width=\hsize]{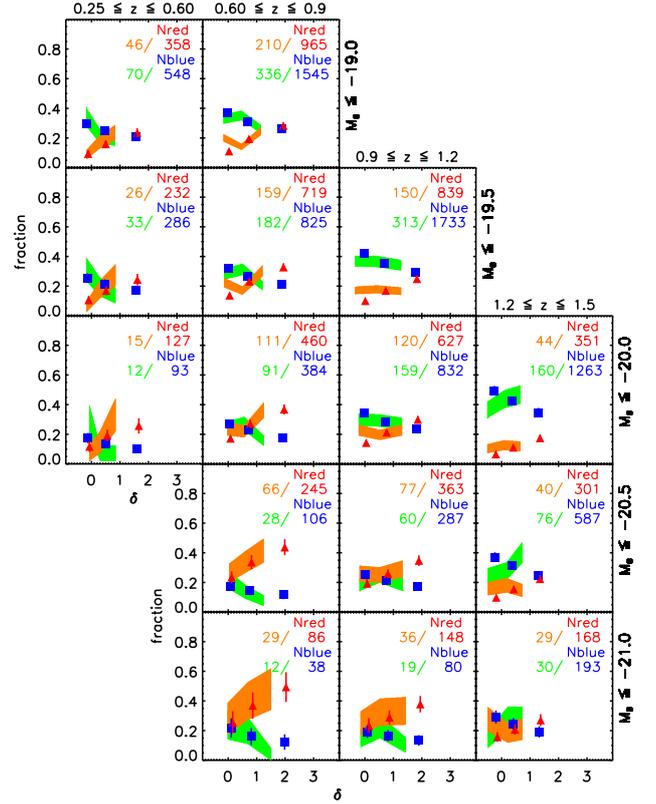}
\caption{Like in Fig.~\ref{col_den_2h}, but for \cmo. }
\label{col_den_FULL} 
\end{figure}

Fig.~\ref{col_den_2h} shows the fraction of red ($f_{red}$) and blue ($f_{blue}$)
galaxies as a function of the density contrast $\delta$, for different redshift
bins and for different luminosity thresholds. Red and blue symbols are for
\omo, while green and orange shaded areas show the error contours for the
VVDS. Fig.~\ref{col_den_FULL} shows the corresponding quantities for \cmo. The
results from \omos have larger error bars due to larger Poisson error, but the
average is very close to that of \cmo. In each mock, we computed $f_{red}$ and
$f_{blue}$ as a function of $\delta$ in three equipopulated density bins (choosing
the median $\delta$ value as representative for the bin). The points in
Fig.~\ref{col_den_2h} and \ref{col_den_FULL} are the average of the mocks, on
both x- and y-axis. Vertical error bars are the sum in quadrature of the $rms$
around the mean of the 23 mocks and the typical error (Poisson) of one single
mock. In this way, we simultaneously account for cosmic variance among
different realizations, and for the typical Poisson noise in a single
realization.

\subsubsection{The effects of a $<$100\% sampling rate on the
  colour-density relation}\label{col_den_full_vvdslike}

We find that the average colour-density relation in \omos is very similar to
the one found in \cmo. This means that, on average, the VVDS observational
strategy does not significantly alter the observed environmental trends for
galaxy colours, on a $\sim5$h$^{-1}$Mpc scale.  This does not exclude the
possibility that there may exist at least one mock in which the colour-density
relation is strongly affected by the VVDS observational strategy. In order to
address this question, we performed a linear fit of $f_{red}$ as a function of
$\delta$ in each luminosity and redshift bin considered, and compared the slope
of this linear fit in \cmos and \omo. We did the same calculation for $f_{blue}$.

\begin{figure} \centering
  \includegraphics[width=9cm]{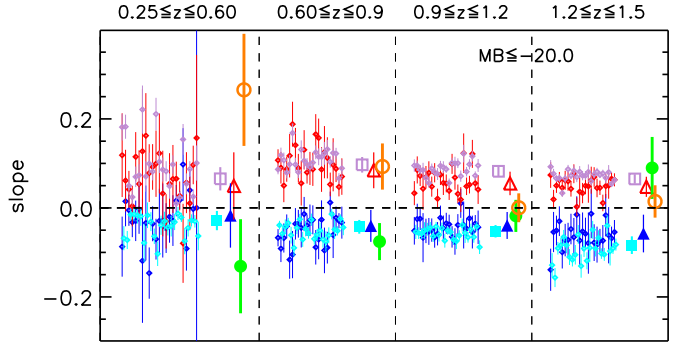}
  \caption{Slopes of the colour-density relation linear fits, for the
    23 \cmos and the 23 \omo, for galaxies with $M_B\leq -20$, in the
    four redshift bins (label on the top), like in the third row of
    Fig.~\ref{col_den_2h} and Fig.~\ref{col_den_FULL}. Red and blue
    small diamonds: red and blue galaxies in each \uomo; light purple
    and cyan small diamonds: red and blue galaxies in each \ucmo. Red
    and blue big triangles: mean values of red and blue small
    diamonds, with the vertical error bar being the rms among all the
    23 mocks. Light purple and cyan big squares: mean values of light
    purple and cyan small diamonds, with their rms.  Orange and green
    circles: slope, and its error, for VVDS observed data (from C06),
    for red and blue galaxies respectively. }
\label{col_den_diff} 
\end{figure}

In Fig.~\ref{col_den_diff}, we show the slopes
obtained for each mock, in the four redshift bins considered and for
galaxies with $M_B \leq -20.$ (i.e., the third row of
Fig.~\ref{col_den_2h}). We plot the slope values for both \cmos and
\omo. The x-axis value is arbitrary, but for each couple of \cmos and
\omos extracted from the same light cone we use the same x-axis value
for $f_{red}$ and $f_{blue}$.  Fig.~\ref{col_den_diff} shows
that a positive slope in \cmos never becomes negative in the \omo, or
vice-versa, but for a couple of exceptions at $0.25<z<0.6$ which may
be due to low number statistics. The figure also show the average
slopes for \cmos and \omo, together with the slopes from the VVDS
sample. 

This Figure shows that going from \cmos to \omo, the
colour-density relation becomes slightly shallower, but does not
disappear nor reverse.  This suggests that the trends found in C06 (the
flattening and possibly the reversal of the colour-density relation
going to high redshift) are  not due to biases in the VVDS observational
strategy (uneven and $<100$\% sampling rate, field shape, etc.).

\subsubsection{Observed data versus mocks}\label{col_den_mocks_real}

In the previous Section, we showed that the VVDS strategy does not
significantly alter the colour-density relation potentially observable
in a survey with 100\% sampling rate. Now we analyse the colour-density
relation in the VVDS observed sample and the one found in mocks, with the aim to
contrast galaxy evolution in simulations and in the real Universe.  

Some interesting trends are visible in Fig.~\ref{col_den_2h} and
\ref{col_den_FULL}. First, the density contrast in mocks spans a
larger range than observed data, extending towards higher densities.  This
mirrors the stronger clustering found in mocks and discussed in Paper I and in
Sec.~\ref{mocks_sec}. Second, error bars for \cmos are 
smaller than those of VVDS, although they include the cosmic variance
among the 23 mocks. This is because the Poisson noise is much lower,
as can be seen comparing the number of galaxies
shown in the labels of Fig.~\ref{col_den_FULL}. In contrast, error
bars for \omos are larger than those of VVDS, because their Poisson
noise is similar but they include also cosmic variance.  These
trends are similar for all redshift bins and luminosity thresholds considered.

At a fixed luminosity threshold, the colour-density relation in VVDS
is steeper than in mocks at $0.25<z<0.6$, very similar to that in
mocks at $0.6<z<0.9$, and shallower than in mocks at $z>0.9$. In
particular, the colour-density relation in the VVDS data is almost
flat at $0.9<z<1.2$, and it seems to be inverted at $z>1.2$ (i.e., at
these high $z$ blue galaxies reside preferentially in high density
regions). In contrast, the mock colour-density relation varies only
weakly as a function of redshift with no significant flattening, and
definitely no inversion at higher redshift. We analysed in details the
colour-density relation in the couple of \ucmo/\uomos with the lowest
number counts. As for the density distribution shown in
Fig.~\ref{delta_distrib}, this mock shows a colour-density relation
very close to the average of the 23 mocks. We will discuss the
implications of these general trends in Sect.~\ref{discussion}.

\begin{figure*} 
\centering
  \includegraphics[width=13cm]{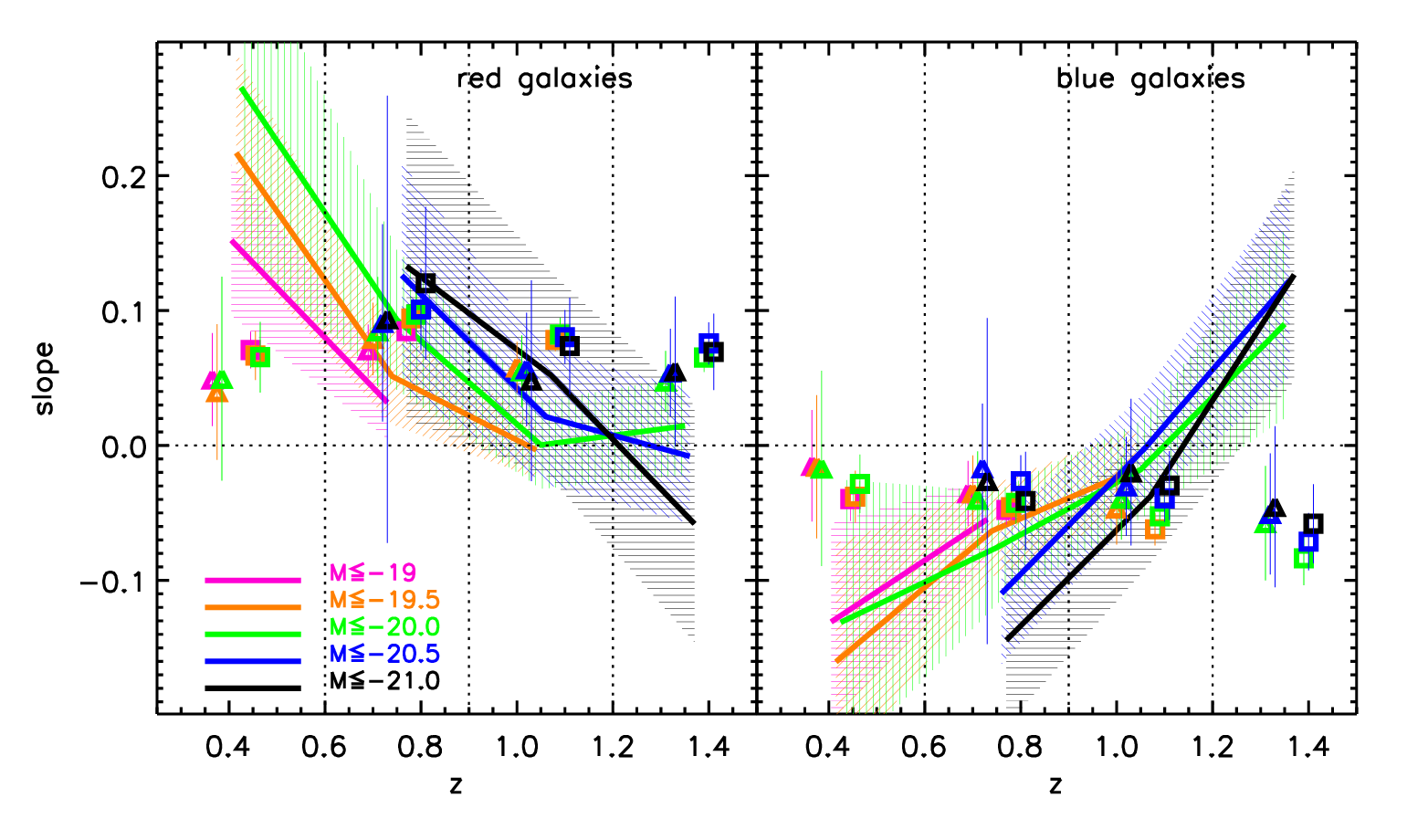}
  \caption{Slopes of the linear fits of the colour-density relation, in each panel of
    Fig.~\ref{col_den_2h} and Fig.~\ref{col_den_FULL}, for $f_{red}$ on the
    left panel and $f_{blue}$ in the right panel. Lines and shaded areas:
    VVDS observed data with error. Triangles: \omo. Squares: \cmo.
    Different colours of lines and symbols are for the different
    luminosity thresholds, as indicated in the labels. In each
    redshift bin, delimited by vertical dashed lines, all the points
    should be considered at the same (central) x-value, but they are
    shifted for clarity. } 
\label{allslopes} 
\end{figure*}

Fig.~\ref{allslopes} shows the slopes obtained from fitting the
colour-density relation in the mocks and compares them with those
obtained from the VVDS data.  Although the error bars for the VVDS
data are large, the evolution of the colour-density relation with
redshift is clear. In contrast, no evolution is found in \omo, nor in
\cmo. Moreover, while in the VVDS data there is a trend for steeper
colour-density relation for brighter galaxies, this trend is much less
clear (if any) in mocks.  We note that the dependence of the
colour-density relation on galaxy luminosity is a controversial issue
in the literature. It has been found in C06, but not in
\cite{cooper2007_col_env}, who studied the colour-density relation in
the DEEP2 survey \citep{Davis2003}, selected with a slightly brighter
flux limit than the VVDS. In \cite{cucciati2010_zCOSMOS}, we did not
find any luminosity dependency in the zCOSMOS Bright survey
\citep{lilly2009_zcosmos}, which is much brighter ($I_{AB}\leq22.5$)
than the sample used in C06.

The comparison between the slopes measured in the mocks and those
measured from data may suffer from the fact that the density
distribution in mocks extends to higher densities than
observed. Because of this, in almost each panel of
Fig.~\ref{col_den_2h} and \ref{col_den_FULL} the point at highest
density has a larger x-value in mocks than in the VVDS, even if $f_{red}$
or $f_{blue}$ are very similar. Thus the slope may be flatter in mocks than
in the VVDS, only because of the different density distribution. To do
a more fair comparison, we computed the fractional increment of $f_{red}$
and $f_{blue}$ from the lowest to the highest density bin in each panel of
Fig.~\ref{col_den_2h} and \ref{col_den_FULL}. The results, for both
observations and mocks, are very similar to those of
Fig.~\ref{allslopes}, and confirm that there is no significant
flattening of the colour-density relation in the mocks.

The results discussed above confirm that, at least in the
  semi-analytical model used in our study, the environment shapes
  galaxy evolution (as a function of $z$) in a different way than we
  observe in the real Universe.

\section{Why is there a colour-density relation? }\label{sec_halos}

Different methods to parameterise the local density around galaxies
have been used in the literature for different galaxy samples. Often
it is the survey strategy itself that dictates the optimal (less
biased) environment parameterisation for each specific data set. It is
not yet clear which is the `physical meaning' of these
parameterisations, e.g. how the density field compares to the
underlying dark matter distribution, and in particular how the
estimated `density' relates to the mass of the DM halos in which
galaxies reside.

Some recent studies \citep{Haas_etal_2012,Muldrew_etal_2012} have
focused on a comparison between different environmental definition and
the information on the parent DM halo mass. These studies have
considered results at $z\sim0$, and have not entered the details of
different observational strategies and/or selections. Given the
evolution of the mean density and of the mass growth of structures, it
would be very interesting to extend this detailed analysis to higher
redshift. It should be noted, however, that results might well depend
on the details of each particular survey and, as such, they are
difficult to generalise.  

\begin{figure}
  \centering
  \includegraphics[width=9cm]{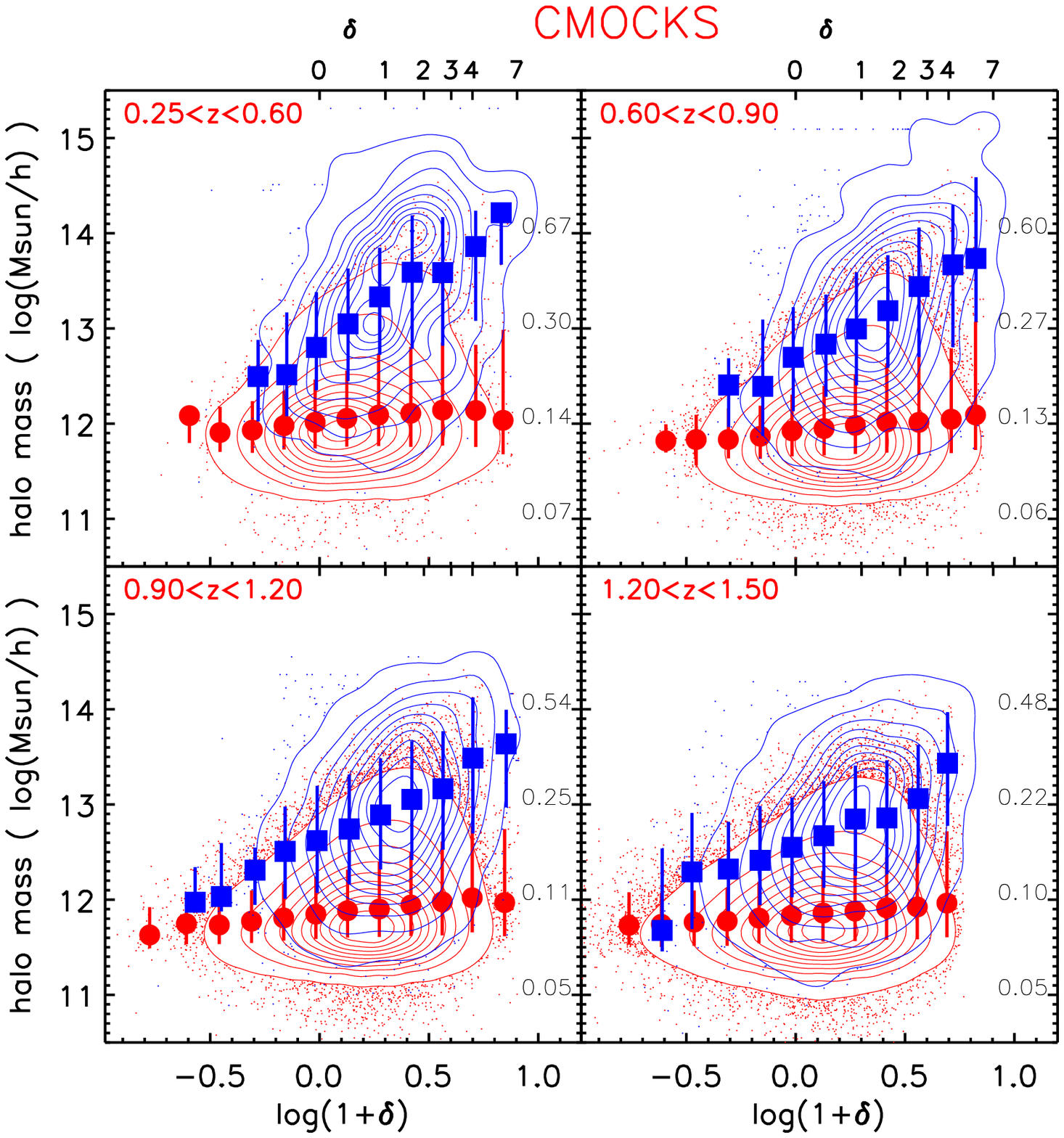}
  \includegraphics[width=9cm]{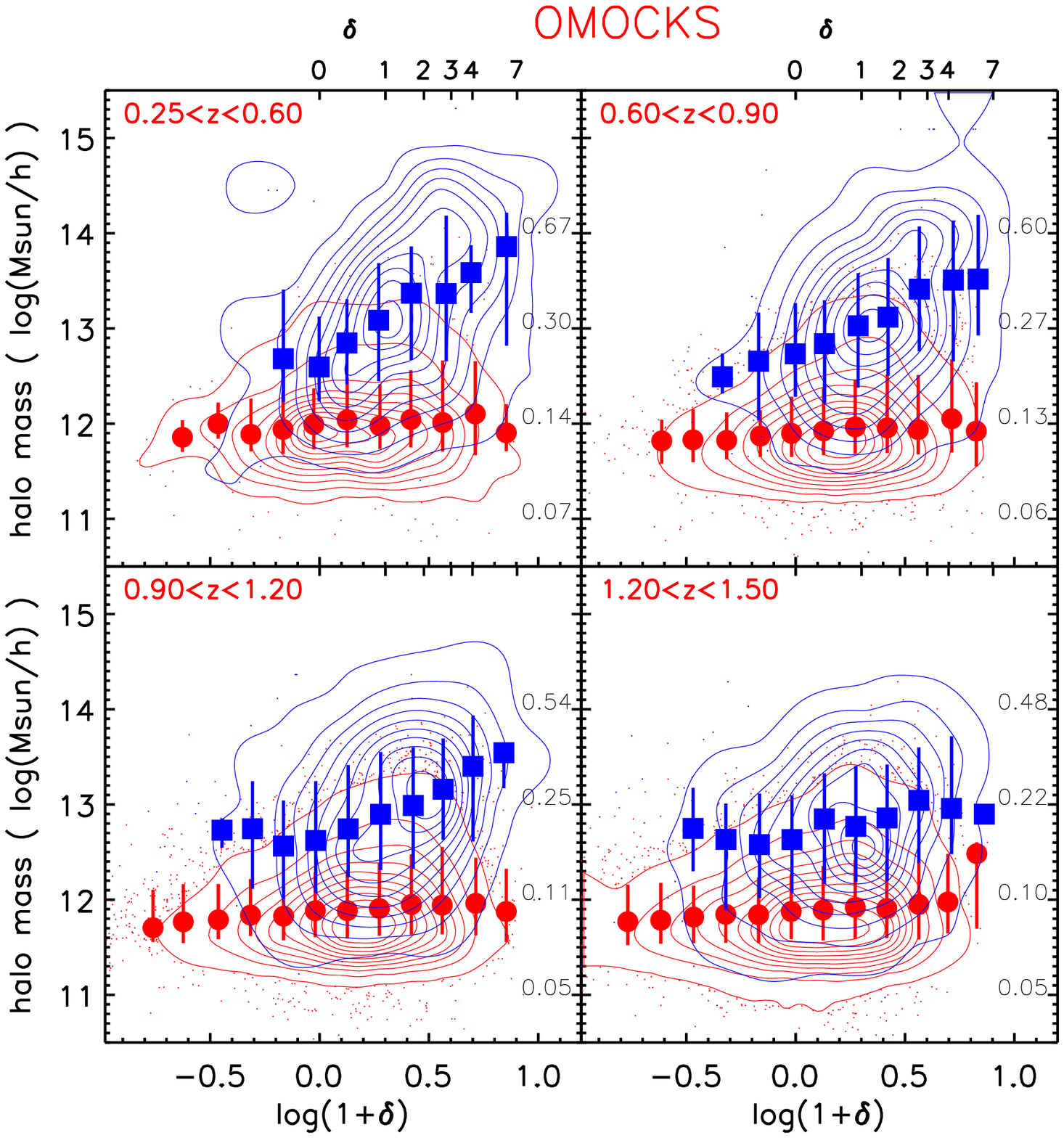}
  \caption{Total halo mass as a function of the local density contrast
    computed in this work, in different redshift bins (as in the
    labels), for galaxies with $M_B\leq-20$. The values of the density contrast are indicated as
    $log(1+\delta)$ in the bottom x-axis and as $\delta$ in the top
    x-axis. The halo is the FOF halo in which the galaxy reside (the
    halo of the group/cluster of which the galaxy is member). Four top
    panels: \cmo. Four bottom panels: \omo.  Red circles and contours
    refer to central galaxies, blue squares and contours to satellite
    galaxies. Each panel includes all the 23 mocks. Lines represent
    the isodensity contours of all the galaxies in the panel. Not to
    crowd the plot, single galaxies are plotted as small dots only
    outside the lowest density contour. Filled red circles (/blue
    squares) are the median values, in density bins, for central
    (/satellite) galaxies. The error bars represents the 16\% and 84\%
    of the halo mass distribution in each density bin. On the right
    vertical axis of each panel, we show the typical virial radius (in
    $h^{-1}$Mpc) corresponding to the halo mass on the left vertical
    axis. }
  \label{halo_vs_dens} 
\end{figure}

We have taken advantage of the available MILLENNIUM mocks to understand
the origin of the colour-density relation as observed in VVDS data. To
do this, we explored the relationship existing between the estimated
density contrast and the mass of the DM haloes in which galaxies
reside. For each galaxy in our mocks, we retrieved its parent DM halo
using the public database built for the Millennium Simulation
\citep{lemson2006_database}.  Here, a `DM halo' corresponds to a halo
identified in the Millennium Simulation using a friends-of-friends
algorithm, with a linking length of 0.2 in units of the mean
interparticle separation. Our results are shown in
Fig.~\ref{halo_vs_dens}, for galaxies with $M_B\leq-20$. We
repeated our analysis for both \cmos and \omos (top and bottom panels
of the figure), in order to test the influence of the VVDS
observational strategy.  In each panel, we distinguish central
galaxies from satellite galaxies. On the right vertical axes of
Fig.~\ref{halo_vs_dens}, we indicate the typical radius (in
$h^{-1}$Mpc) of haloes with mass given by the corresponding value on
the left axis\footnote{The virial mass is computed using the simulated
  particles, as the mass enclosed within a sphere that corresponds to
  an overdensity of 200 times the critical density.  The virial radius
  is computed from the virial mass, through scaling laws based on
  simulation results and the virial theorem.}. This shows that the
scale on which $\delta$ is computed is much larger than the size of
the structures in which galaxies reside.

 Fig.~\ref{halo_vs_dens} shows a very general trend: galaxies
belonging to massive halos
  (mass $\gtrsim 10^{13}M_{\bigodot}/h$) reside only in over-dense
  regions on a $5 h^{-1} Mpc$ scale, even if the virial radius of such
  halos is much smaller than the filtering scale used. This happens
  because the density in these regions is boosted to high values by
  the large number of galaxies residing in these massive haloes, and
  is not affected significantly by the large scale structure around
  them.  In contrast, lower
  mass halos span the entire density range, because their density
  within the virial radius is not very high, so the density on larger
  scales depends also on the surrounding structures.

  From Fig.~\ref{halo_vs_dens}, it is clear why we should expect a
  colour-density relation on a $5 h^{-1} Mpc$ scale. There is an
  increase of the parent halo mass with increasing $\delta$ for
  satellite galaxies, although with a large scatter. This relation is
  driven by the large number of satellites in the most massive
  halos. The Figure shows that, if a satellite has a high measured
  $\delta$ on the scale considered, it more probably belongs to a
  massive cluster than to a low mass halo. If we assume that clusters
  have higher fraction of red satellites than groups \citep[see
  e.g. ][]{delucia2012_histories}, we expect a colour-density relation
  for satellite galaxies.  

For central galaxies, the correlation between halo mass and $\delta$
on $5 h^{-1} Mpc$ is weak, but there is still a trend for central
galaxies in highest density regions being scattered to higher mass
haloes. This happens because the most massive haloes tend to be
clustered. Then, if we assume that central galaxies in more massive
haloes are redder than central galaxies of lower mass haloes, we would
expect to have a higher fraction of red galaxies in higher density
regions.

  In Fig.~\ref{col_den_2h} and \ref{col_den_FULL}, we showed that we
  do find a colour-density relation for model galaxies. We verified
  that this relation is maintained when considering only centrals or
  only satellites galaxies. The colour-density relations for the two
  populations are similar, and in agreement with the global
  colour-density relation.

As redshift increases, the VVDS flux limits select brighter and
brighter galaxies. Thus, the fraction of observed satellites (that dominate
the intermediate to faint end of the LF) decreases, so that the
correlation between density and halo mass becomes less significant
also for satellites.  This is also evident comparing \cmos with \omo:
the VVDS observational strategy applied in \omos reduces the fraction
of observed satellites in massive clusters, flattening the relation
between halo mass and $\delta$. This happens because the size of the
slits in the spectrograph prevents us from targeting, in one single
pass, galaxies with small projected distances. The VVDS multi-pass
strategy alleviates this problem but small-scale very dense regions
(like the central regions of galaxy clusters) are still undersampled
with respect to regions that are less crowded. At the highest redshift
considered, where the effect of the low sampling rate sums up with the
effect of the flux limit, we do not see any clear correlation, even
for satellite galaxies.

In a survey like the VVDS, given its flux limit, the fraction of
satellites galaxies ($f_{sat}$) is small, and it decreases with
$z$.  In particular, for galaxies with
  $M_B\leq-20$, it varies from $\sim15$\% to $\sim8$\% going from the
  lowest to the highest redshift explored. So, we expect that central
  galaxies have an important role in shaping the observed 
  colour-density relation. We will come back to this in
  Sec.~\ref{discussion}.

\section{Discussion}\label{discussion}

As shown in previous studies and above, the Munich semi-analytical
model does not reproduce some of the galaxy properties measured from
the VVDS. In particular, in the model there is an excess of sources
for $I_{AB}<24$, mainly due to an excess of faint red galaxies. The
model also exhibits a slight excess of bright blue galaxies, and the
galaxy clustering is overpredicted. In this paper, we have studied how
in this model the 3D galaxy distribution affects galaxy colours up to
$z=1.5$.  The aim of our analysis is to understand how environment
affects galaxy evolution modulating internal physical processes, and
to what extent this is reproduced by the model used in this study.

Our analysis demonstrates that the colour-density relation, computed
on a 5$h^{-1}$Mpc scale, does not evolve significantly with redshift
in the Munich semi-analytical model, at least from $z=0.25$ to
$z=1.5$. This relation does not evolve even if we consider
  centrals and satellites separately.  In contrast, significant
evolution has been observed in different samples and on different
scales (e.g.  C06, \citealp{cooper2007_col_env},
\citealp{cucciati2010_zCOSMOS}), and C06 even observed a possible
inversion in such relation at $z\gtrsim 1.2$.

Given the striking difference between observational results and model
predictions about the evolution of the colour-density, some questions arise:
within a cosmological framework, do we expect an epoch when the
colour-density relation is inverted? On which scale should such an
inversion be measured? And are the physical processes responsible for
such a relation included in the model that we have considered in this
study?

Qualitatively, we expect a reversal of the colour-density
relation, at an early epoch and on the scale of galaxy groups, due to
starburst events in gas-rich mergers. In this scenario, two young and
gas-rich galaxies will evolve differently if one remains in isolation
(passive evolution) and the other merges with another gas-rich galaxy
(this will trigger a starburst episode whose intensity depends on the
mass ratio and on the amount of gas available). During the merger, and
for some time after it has been completed, the fraction of bright
star-forming galaxies should be higher in high densities than in low
densities. This simple scenario is complicated by the fact that lower
density environments will also contain young star forming galaxies,
although these might be fainter than the VVDS flux limit. Another
complication is due to dust attenuation: a large fraction of the
galaxies that are classified as red might be forming stars at some
significant rate, and this fraction might evolve as a function of
cosmic epoch and/or luminosity and
environment. For example,  \citet{delucia2012_histories} find that the fraction
of red star forming galaxies decreases with increasing mass and with
decreasing distance from the cluster center. Moreover, it has been
observed that dust attenuation clearly evolves with redshift
\citep[see e.g. ][]{cucciati2012_sfrd}.

A similar scenario, based on starburst events in wet mergers, has been
proposed by \cite{elbaz2007}, who find an inversion of the star
formation - density relation comparing the GOODS fields at high
redshift with SDSS data. In particular, they find the mean star
formation rate to be larger in high densities at $z\sim1$, on a
$1.5h^{-1}$Mpc scales. Their galaxy selection is similar to the one
adopted in C06 at the same redshift ($M_B\leq-20$).  In their sample,
these galaxies are often located in correspondence of local density
enhancement on the scale of clusters/groups
($\sim1h^{-1}$Mpc). \citet{elbaz2007} also measure the SF-density
relation in SAMs (the model by \citealp{croton2006} applied on the
Millennium Run by \citealp{kitzbichler2007}), applying the same
methods as in their observational sample, and they did not find any
reversal at $z\sim1$, but a mild reversal of the SF-density relation
at $z\sim2$. \cite{wang2007_model} carried out a similar analysis
focusing on the relation between the average D4000$\AA$ break and the
local density computed on a $\sim2h^{-1}$Mpc scale. In agreement with
what discussed above, they find that in the Munich semi-analytical
model there is no significant weakening of the D4000$\AA$-density
relation up to redshift $\sim 3$.

It is interesting that in C06 we found a possible reversal of the
colour-density relation on $\sim5h^{-1}$Mpc scales, that is much larger than
those of galaxy groups. A few previous studies have argued that large scale
environmental trends are the residual of the trends observed on much smaller
scales (\citealp{Kauffmann2004,blanton2006, cucciati2010_zCOSMOS}, and see also
\citealp{wilman2010_multiscale}).  We can not establish if this is the case
also in C06, as the scale used to compute the density contrast was the smallest
allowed by the VVDS sampling rate to guarantee a reliable reconstruction of the
density field (see the tests in C06). In addition, the $\sim5h^{-1}$Mpc scale
should be compared to the typical scales of processes happening `around'
clusters, like for instance the infall of galaxies and groups onto larger
structures. One may also expect that such scale varies with redshift, as
structures grow with time.

 It
  is interesting that in the Munich semi-analytical model we find the
  same (not-evolving) colour-density relation for satellites and
  central galaxies separately (see Sec.~\ref{sec_halos}). We should
  consider that some physical processes that might be at play
in determining the evolution and the inversion of the colour-density
relation are modelled in a rather crude way. For example, starbursts
triggered by mergers are `instantaneous' (i.e. their time-scale
corresponds to the integration time-scale adopted in the model), and
the model does not account for star formation episodes that might be
triggered during fly-by. In addition, in this model (like in most of
the recently published ones) there is a significant excess of red
(passive) galaxies with respect to observations. In particular, the
model galaxies in the region of infall around clusters are redder than
in the real Universe, because most of the galaxies in a group that is
infalling in a cluster will be already too red in the model.

\begin{figure}
  \centering
  \includegraphics[width=9cm]{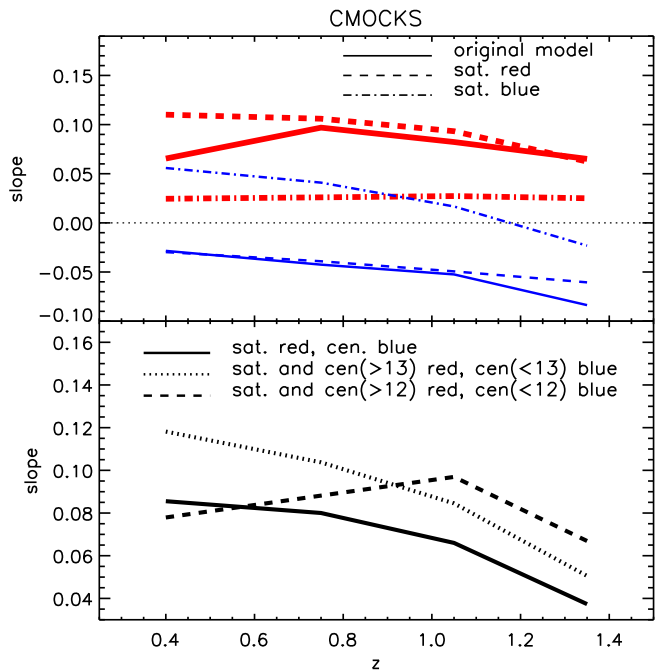}
  \caption{{\it Top panel}. Slope of the linear fit of the $f_{red}$-density
  relation (thick red lines) and of the $f_{blue}$-density
  relation (thin blue lines) as a function of redshift in \cmo. 
  Only galaxies with$M_B\leq-20$ are considered. Solid line: slopes as 
  found in the model, the same as in Fig.~\ref{allslopes}. Dashed 
  and dot-dashed lines: slopes for two 
  different assumptions about the colour distributions of satellite
  galaxies. We assume that centrals galaxies have the same colour as in 
  the model, but now we assume that satellites are either all red 
  (dashed line) or all blue (dot-dashed line).
  {\it Bottom panel}. Slope of the linear fit of the $f_{red}$-density
  relation as a function of redshift in \cmo, for three different assumptions
  about the colour distributions of central and satellite
  galaxies. Solid line: all satellites are red, and all centrals are
  blue. Dotted line: all satellites and centrals residing in DM halos
  with virial mass $\geq10^{13}M_{\odot}/h$ are red, while all other centrals
  are blue. Dashed line: like the dotted line, but in this case the
  red centrals are those residing in DM halos with virial mass
  $\geq10^{12}M_{\odot}/h$. For this plot we use the 23 \cmos
  together, and we keep $f_{sat}$ increasing with $\delta$ and
  decreasing with $z$ as we find in the mocks.}
\label{toy_model}
\end{figure}

Since the model does not match the expected evolution of the
colour-density relation as a function of redshift, we can modify the
colour distribution in the model by making simple assumptions, and see
how the colour-density relation changes. For example, we verified 
what happens if we assume that the colour of central galaxies is correctly 
reproduced by the model, and that satellites are either all red or all
blue. The results of this test are shown in the top panel of 
Fig.~\ref{toy_model}. Since $f_{sat}$ increases with density, at all $z$, if
satellites are all red the increase of $f_{red}$
as a function of density would be even steeper than what
found in the model. So this would increase the disagreement between
data and models. If we assume that satellites are all blue, which is
unrealistic, we find that the $f_{red}$-$\delta$ relation flattens at
all redshifts, with the flattening being a bit more significant at low
redshift. This is because $f_{sat}$ decreases with increasing redshift
(see Fig.~\ref{halo_vs_dens}). This is, however, not enough to have a
flat relation at $z>1.2$, as observed for VVDS. In addition, assuming
all satellites are blue, would give an almost flat $f_{blue}$-$\delta$
relation at $z>1.2$ but would invert the relation at $z<0.9$, contrary
to what is observed.

This suggests that an inaccurate modelling of the physics of satellite
galaxies cannot be solely responsible for the disagreement we find
between models and data, and that such a discrepancy also signals the
need for a revised treatment of central galaxies.

In order to gain insight on this, let us assume that the model
predicts the correct increase (decrease) of $f_{sat}$ as a function of
density (redshift), and that all satellites are red while all centrals
are blue. Since $f_{sat}$ decreases with increasing redshift, this
would make the $f_{red}$-$\delta$ relation flatten with increasing
redshift, as shown in the bottom panel of Fig.~\ref{toy_model}. The 
figure shows that a
flattening would be observed also in the case we assume all satellites
and central galaxies of haloes more massive than $10^{13}M_{\odot}/h$
(dotted line)  are red, and that all other centrals are blue.
On the contrary, if we assume to be red all satellites and all central 
galaxies 
of haloes more massive than $10^{12}M_{\odot}/h$
(dashed line),   the evolution of the slope of the $f_{red}$-$\delta$
is not monotonic.

  The toy-model considered is simplistic, because for instance we
  consider all satellites to be red, and we do not include any
  redshift evolution of the colour-mass relation for
  centrals.  Anyway, this simple
  exercise tells us that, for a survey like VVDS, the role of central
  galaxies in the evolution of the colour-density relation is very
  important, and that also the modelling of these galaxies needs to be
  revised in the model used in this study.

Further work is therefore needed in order to understand what are the
physical processes driving the flattening (and possibly inversion) of
the colour-density relation in real observations, and how to include
them properly in the context of hierarchical galaxy formation models.
An interesting approach would be to use the accretion histories of
model galaxies in different density bins, as done in
\cite{delucia2012_histories}.  If galaxies in low and high density
regions have spent different fractions of their life-time in high
density environments, we expect them to have different colours. Using
this approach, it will be possible to compare, and possibly to link,
the formation histories of galaxies at low and high redshift, and to
shed light on the evolution of the colour-density relation. We will
address this issue in future work.


\section{Summary and Conclusions}

In this paper we used galaxy mock catalogues
constructed from a semi-analytic model coupled to a large cosmological
simulation, to compare the observed colour-density analysis up to $z\sim1.5$
presented in \cite{cucciati2006} with model predictions, reproducing
carefully in the mocks the observational selection and strategy
adopted for the VVDS Deep sample.

In particular, we used 23 galaxy mock catalogues, with the same flux limits
adopted in the VVDS-Deep survey (\cmo). The mocks are extracted from the
Millennium Simulation, with applied the semi-analytical model by
\cite{delucia_blaizot2007}. From each of these mocks, a sub-sample of galaxies
has been extracted (\omo) mimicking the VVDS survey observational strategy
(sampling rate, slit positioning, etc.). We then computed the galaxy luminosity
function and the colour-density relation in the mocks, using the same methods
employed for the VVDS survey. Our results can be briefly summarised as follows:

\begin{itemize}

\item[1)] the mocks based on the Munich semi-analytical model contain on
  average $\sim10$\% more galaxies than the VVDS sample. This is consistent
  with results obtained in Paper I.

\item[2)] The rest-frame B-band LF in mocks is in rough agreement with
the one derived from the observations,
  but it has a slightly steeper slope, especially at $0.2<z<0.8$. This is
  expected, given the excess of galaxies in the mocks in this redshift
  range. We have also computed the LF for early and late type galaxies
  separately, and shown that the model over-predict the number densities of
  faint early type galaxies and that of bright late type
  galaxies.

\item[3)] The density distribution computed with the same method as in the VVDS
  has more prominent tails towards higher densities in the mocks, at
  any redshift and luminosity explored. This is not due to the larger number
  counts in mocks, but it is related to an intrinsically different galaxy
  distribution, which also reflects in a stronger clustering signal in mocks
  (see Paper I).

\item[4)] The colour-density relation in \omos is in very good agreement with
  the one in \cmos (being only more noisy and slightly less
  significant in \omo). This enhances the confidence  that the
  evolutionary trend of the colour-density relation in the VVDS-Deep
  survey is not caused by any severe bias introduced by the survey
  observational strategy.

\item[5)] The colour-density relation in mocks does not evolve significantly
  from $z=0.25$ to $z=1.5$, in contrast with a significant evolution measured
  in the data. In particular, we find no flattening (or inversion) of the
  colour-density relation at higher redshift, at least up to the redshift
  explored by the VVDS data and on the same scale.

\item[6)] Given the relation between the measured density
 contrast and the virial mass of the halos where galaxies reside, we
 do expect and find a colour-density relation both for central and
 satellites galaxies. Both are very similar to the relation found for
 the global population. We argue that the lack of evolution in the
 colour-density relation in mocks cannot be due only to inaccurate
 prescriptions for the evolution of satellites galaxies, and that also the
 treatment of the central galaxies has to be revised.

\end{itemize}

A reversal of the colour-density relation is expected in a scenario in
which wet mergers of young galaxies trigger an enhancement of star
formation in the interacting galaxies. In this scenario, a reversal of
the colour-density relation should be observed on the scale of galaxy
groups. A reversal of the star formation-density relation has been
observed on such scales at $z\sim1$ \citep{elbaz2007}. However, it is
not straightforward to correlate galaxy SF with colours, because of
dust reddening. So it is not clear that what we measure on a
$5h^{-1}$Mpc scale is a mirror of what is happening on smaller
scales. The lack of evolution of the colour-density relation in the
model, on a $5h^{-1}$Mpc and at least up to $z=1.5$, suggests that the
evolution has happened at $z>1.5$,  and/or that model
galaxy colour is affected by environment on scales much smaller than
5$h^{-1}$Mpc, with no corresponding trends on larger scales.

The disagreement between the evolution of the colour-density relation
in mocks and in observed data deserves further investigation as it can
clarify what are the physical processes driving the observed
flattening (and inversion), and to what extent these physical
processes are included in recent models of galaxy formation. It
would be also important to disentangle the role of central and
satellites galaxies. In this work we focused on mocks reproducing the
VVDS observational strategy. With these mocks it is not possible to go
further in this analysis, because the flux limits prevent us from
having enough galaxies  to study in details the small-scale
environment at the highest redshift explored, even if we had a 100\%
sampling rate.  With mock catalogues it is possible to go beyond the
luminosity and redshift range explored by the VVDS, but currently
there would not be a suitable counter-part among the available data
sets. For example, \cite{knobel2012_groups20K} successfully separate
central and satellites galaxies in the zCOSMOS group catalogue, but
this catalogue reaches only $z=1$, and the flux limit ($I_{AB}\leq
22.5$) is brighter than the one in the
VVDS. \cite{gerke2012_groupsDEEP2} use data from the DEEP2 survey to
compute a group catalogue that extends at $z>1$, but they do not
distinguish between central and satellite galaxies. Moreover, the
DEEP2 flux limit is very similar to the VVDS one, so at $z>1$ the
study of satellite galaxies would be difficult. Therefore, deeper
spectroscopic surveys, with higher sampling rate, would be needed in
order to shed light on the physical processes establishing the
observed relation between colour and density.


\begin{acknowledgements}

We would like to thank the referee for helpful suggestions that
improved the paper.  The Millennium Simulation databases used in this
paper and the web application providing online access to them were
constructed as part of the activities of the German Astrophysical
Virtual Observatory. We are grateful to Gerard Lemson for setting up
an internal data base that greatly facilitated the exchange of data
and information needed to carry out this project, and for his
continuous help with the database.  OC thanks the INAF-Fellowship
program for support. GDL acknowledges financial support from the
European Research Council under the European Community's Seventh
Framework Programme (FP7/2007-2013)/ERC grant agreement
n. 202781. Part of this work was supported by PRIN INAF 2010 `From the
dawn of galaxy formation to the peak of the mass assembly'. The
results presented in C06 have been obtained within the framework of
the VVDS consortium, and we thank the entire VVDS team for its work.

\end{acknowledgements}

\bibliographystyle{aa}
\bibliography{biblio}

\begin{thebibliography}{66}
\expandafter\ifx\csname natexlab\endcsname\relax\def\natexlab#1{#1}\fi

\bibitem[{{Abbas} \& {Sheth}(2005)}]{abbas2005}
{Abbas}, U. \& {Sheth}, R.~K. 2005, \mnras, 364, 1327

\bibitem[{{Bielby} {et~al.}(2011){Bielby}, {Hudelot}, {McCracken}, {Ilbert},
  {Daddi}, {Le F{\`e}vre}, {Gonzalez-Perez}, {Kneib}, {Marmo}, {Mellier},
  {Salvato}, {Sanders}, \& {Willott}}]{bielby2011_WIRDS}
{Bielby}, R., {Hudelot}, P., {McCracken}, H.~J., {et~al.} 2011, ArXiv:1111.6997

\bibitem[{{Blaizot} {et~al.}(2005){Blaizot}, {Wadadekar}, {Guiderdoni},
  {Colombi}, {Bertin}, {Bouchet}, {Devriendt}, \& {Hatton}}]{Blaizot2005}
{Blaizot}, J., {Wadadekar}, Y., {Guiderdoni}, B., {et~al.} 2005, \mnras, 360,
  159

\bibitem[{{Blanton} {et~al.}(2006){Blanton}, {Eisenstein}, {Hogg}, \&
  {Zehavi}}]{blanton2006}
{Blanton}, M.~R., {Eisenstein}, D., {Hogg}, D.~W., \& {Zehavi}, I. 2006, \apj,
  645, 977

\bibitem[{{Bruzual} \& {Charlot}(2003)}]{BC03}
{Bruzual}, G. \& {Charlot}, S. 2003, \mnras, 344, 1000

\bibitem[{{Bruzual A.} \& {Charlot}(1993)}]{BC1993}
{Bruzual A.}, G. \& {Charlot}, S. 1993, \apj, 405, 538

\bibitem[{{Chabrier}(2003)}]{chabrier2003_IMF}
{Chabrier}, G. 2003, \pasp, 115, 763

\bibitem[{{Coleman} {et~al.}(1980){Coleman}, {Wu}, \& {Weedman}}]{CWW1980}
{Coleman}, G.~D., {Wu}, C., \& {Weedman}, D.~W. 1980, \apjs, 43, 393

\bibitem[{{Cooper} {et~al.}(2007){Cooper}, {Newman}, {Coil}, {Croton}, {Gerke},
  {Yan}, {Davis}, {Faber}, {Guhathakurta}, {Koo}, {Weiner}, \&
  {Willmer}}]{cooper2007_col_env}
{Cooper}, M.~C., {Newman}, J.~A., {Coil}, A.~L., {et~al.} 2007, \mnras, 376,
  1445

\bibitem[{{Cowie} {et~al.}(1996){Cowie}, {Songaila}, {Hu}, \&
  {Cohen}}]{cowie96}
{Cowie}, L.~L., {Songaila}, A., {Hu}, E.~M., \& {Cohen}, J.~G. 1996, \aj, 112,
  839

\bibitem[{{Croton} {et~al.}(2006){Croton}, {Springel}, {White}, {De Lucia},
  {Frenk}, {Gao}, {Jenkins}, {Kauffmann}, {Navarro}, \& {Yoshida}}]{croton2006}
{Croton}, D.~J., {Springel}, V., {White}, S.~D.~M., {et~al.} 2006, \mnras, 365,
  11

\bibitem[{{Cucciati} {et~al.}(2010){Cucciati}, {Iovino}, {Kova{\v c}},
  {Scodeggio}, {Lilly}, {Bolzonella}, {Bardelli}, {Vergani}, {Tasca}, {Zucca},
  {Zamorani}, {Pozzetti}, {Knobel}, {Oesch}, {Lamareille}, {Caputi},
  {Kampczyk}, {Tresse}, {Maier}, {Carollo}, {Contini}, {Kneib}, {Le F{\`e}vre},
  {Mainieri}, {Renzini}, {Bongiorno}, {Coppa}, {de la Torre}, {de Ravel},
  {Franzetti}, {Garilli}, {Le Borgne}, {Le Brun}, {Mignoli}, {Pell{\`o}},
  {Peng}, {Perez-Montero}, {Ricciardelli}, {Silverman}, {Tanaka}, {Koekemoer},
  {Scoville}, {Abbas}, {Bottini}, {Cappi}, {Cassata}, {Cimatti}, {Guzzo},
  {Leauthaud}, {Maccagni}, {Marinoni}, {McCracken}, {Memeo}, {Meneux},
  {Porciani}, \& {Scaramella}}]{cucciati2010_zCOSMOS}
{Cucciati}, O., {Iovino}, A., {Kova{\v c}}, K., {et~al.} 2010, \aap, 524, A2

\bibitem[{{Cucciati} {et~al.}(2006){Cucciati}, {Iovino}, {Marinoni}, {Ilbert},
  {Bardelli}, {Franzetti}, {Le F{\`e}vre}, {Pollo}, {Zamorani}, {Cappi},
  {Guzzo}, {McCracken}, {Meneux}, {Scaramella}, {Scodeggio}, {Tresse}, {Zucca},
  {Bottini}, {Garilli}, {Le Brun}, {Maccagni}, {Picat}, {Vettolani},
  {Zanichelli}, {Adami}, {Arnaboldi}, {Arnouts}, {Bolzonella}, {Charlot},
  {Ciliegi}, {Contini}, {Foucaud}, {Gavignaud}, {Marano}, {Mazure}, {Merighi},
  {Paltani}, {Pell{\`o}}, {Pozzetti}, {Radovich}, {Bondi}, {Bongiorno},
  {Busarello}, {de La Torre}, {Gregorini}, {Lamareille}, {Mathez}, {Mellier},
  {Merluzzi}, {Ripepi}, {Rizzo}, {Temporin}, \& {Vergani}}]{cucciati2006}
{Cucciati}, O., {Iovino}, A., {Marinoni}, C., {et~al.} 2006, \aap, 458, 39

\bibitem[{{Cucciati} {et~al.}(2012){Cucciati}, {Tresse}, {Ilbert}, {Le
  F{\`e}vre}, {Garilli}, {Le Brun}, {Cassata}, {Franzetti}, {Maccagni},
  {Scodeggio}, {Zucca}, {Zamorani}, {Bardelli}, {Bolzonella}, {Bielby},
  {McCracken}, {Zanichelli}, \& {Vergani}}]{cucciati2012_sfrd}
{Cucciati}, O., {Tresse}, L., {Ilbert}, O., {et~al.} 2012, \aap, 539, A31

\bibitem[{{Davis} {et~al.}(2003){Davis}, {Faber}, {Newman}, {Phillips},
  {Ellis}, {Steidel}, {Conselice}, {Coil}, {Finkbeiner}, {Koo}, {Guhathakurta},
  {Weiner}, {Schiavon}, {Willmer}, {Kaiser}, {Luppino}, {Wirth}, {Connolly},
  {Eisenhardt}, {Cooper}, \& {Gerke}}]{Davis2003}
{Davis}, M., {Faber}, S.~M., {Newman}, J., {et~al.} 2003, in Discoveries and
  Research Prospects from 6- to 10-Meter-Class Telescopes II. Edited by
  Guhathakurta, Puragra. Proceedings of the SPIE, Volume 4834, pp. 161-172
  (2003)., ed. P.~{Guhathakurta}, 161--172

\bibitem[{{de la Torre} {et~al.}(2011){de la Torre}, {Meneux}, {De Lucia},
  {Blaizot}, {Le F{\`e}vre}, {Garilli}, {Cucciati}, {Mellier}, {Pollo},
  {Abbas}, {Bottini}, {Le Brun}, {Maccagni}, {Scodeggio}, {Tresse},
  {Vettolani}, {Zanichelli}, {Adami}, {Arnouts}, {Bardelli}, {Bolzonella},
  {Cappi}, {Charlot}, {Ciliegi}, {Contini}, {Foucaud}, {Franzetti},
  {Gavignaud}, {Guzzo}, {Ilbert}, {Iovino}, {McCracken}, {Marinoni}, {Mazure},
  {Merighi}, {Paltani}, {Pell{\'o}}, {Pozzetti}, {Vergani}, {Zamorani}, \&
  {Zucca}}]{delatorre2011_MILLVVDS}
{de la Torre}, S., {Meneux}, B., {De Lucia}, G., {et~al.} 2011, \aap, 525, A125
  (Paper I)

\bibitem[{{De Lucia} \& {Blaizot}(2007)}]{delucia_blaizot2007}
{De Lucia}, G. \& {Blaizot}, J. 2007, \mnras, 375, 2

\bibitem[{{De Lucia} {et~al.}(2004){De Lucia}, {Kauffmann}, \&
  {White}}]{delucia04_SAM}
{De Lucia}, G., {Kauffmann}, G., \& {White}, S.~D.~M. 2004, \mnras, 349, 1101

\bibitem[{{De Lucia} {et~al.}(2006){De Lucia}, {Springel}, {White}, {Croton},
  \& {Kauffmann}}]{DeLucia_etal_2006}
{De Lucia}, G., {Springel}, V., {White}, S.~D.~M., {Croton}, D., \&
  {Kauffmann}, G. 2006, \mnras, 366, 499

\bibitem[{{De Lucia} {et~al.}(2012){De Lucia}, {Weinmann}, {Poggianti},
  {Arag{\'o}n-Salamanca}, \& {Zaritsky}}]{delucia2012_histories}
{De Lucia}, G., {Weinmann}, S., {Poggianti}, B.~M., {Arag{\'o}n-Salamanca}, A.,
  \& {Zaritsky}, D. 2012, \mnras, 423, 1277

\bibitem[{{Efstathiou} {et~al.}(1988){Efstathiou}, {Ellis}, \&
  {Peterson}}]{efstathiou88_LF_SWML}
{Efstathiou}, G., {Ellis}, R.~S., \& {Peterson}, B.~A. 1988, \mnras, 232, 431

\bibitem[{{Elbaz} {et~al.}(2007){Elbaz}, {Daddi}, {Le Borgne}, {Dickinson},
  {Alexander}, {Chary}, {Starck}, {Brandt}, {Kitzbichler}, {MacDonald},
  {Nonino}, {Popesso}, {Stern}, \& {Vanzella}}]{elbaz2007}
{Elbaz}, D., {Daddi}, E., {Le Borgne}, D., {et~al.} 2007, \aap, 468, 33

\bibitem[{{Farouki} \& {Shapiro}(1981)}]{fk81}
{Farouki}, R. \& {Shapiro}, S.~L. 1981, \apj, 243, 32

\bibitem[{{Fontanot} {et~al.}(2009){Fontanot}, {De Lucia}, {Monaco},
  {Somerville}, \& {Santini}}]{fontanot2009}
{Fontanot}, F., {De Lucia}, G., {Monaco}, P., {Somerville}, R.~S., \&
  {Santini}, P. 2009, \mnras, 397, 1776

\bibitem[{{Franzetti} {et~al.}(2007){Franzetti}, {Scodeggio}, {Garilli},
  {Vergani}, {Maccagni}, {Guzzo}, {Tresse}, {Ilbert}, {Lamareille}, {Contini},
  {Le F{\`e}vre}, {Zamorani}, {Brinchmann}, {Charlot}, {Bottini}, {Le Brun},
  {Picat}, {Scaramella}, {Vettolani}, {Zanichelli}, {Adami}, {Arnouts},
  {Bardelli}, {Bolzonella}, {Cappi}, {Ciliegi}, {Foucaud}, {Gavignaud},
  {Iovino}, {McCracken}, {Marano}, {Marinoni}, {Mazure}, {Meneux}, {Merighi},
  {Paltani}, {Pell{\`o}}, {Pollo}, {Pozzetti}, {Radovich}, {Zucca}, {Cucciati},
  \& {Walcher}}]{franzetti2007}
{Franzetti}, P., {Scodeggio}, M., {Garilli}, B., {et~al.} 2007, \aap, 465, 711

\bibitem[{{Gavazzi} {et~al.}(1996){Gavazzi}, {Pierini}, \&
  {Boselli}}]{gavazzi96}
{Gavazzi}, G., {Pierini}, D., \& {Boselli}, A. 1996, \aap, 312, 397

\bibitem[{{Gerke} {et~al.}(2012){Gerke}, {Newman}, {Davis}, {Coil}, {Cooper},
  {Dutton}, {Faber}, {Guhathakurta}, {Konidaris}, {Koo}, {Lin}, {Noeske},
  {Phillips}, {Rosario}, {Weiner}, {Willmer}, \& {Yan}}]{gerke2012_groupsDEEP2}
{Gerke}, B.~F., {Newman}, J.~A., {Davis}, M., {et~al.} 2012, \apj, 751, 50

\bibitem[{{Gunn} \& {Gott}(1972)}]{gunn_gott1972}
{Gunn}, J.~E. \& {Gott}, J.~R.~I. 1972, \apj, 176, 1

\bibitem[{{Guo} {et~al.}(2012){Guo}, {White}, {Angulo}, {Henriques}, {Lemson},
  {Boylan-Kolchin}, {Thomas}, \& {Short}}]{guo2012_millWMAP7}
{Guo}, Q., {White}, S., {Angulo}, R.~E., {et~al.} 2012, ArXiv:1206.0052

\bibitem[{{Haas} {et~al.}(2012){Haas}, {Schaye}, \&
  {Jeeson-Daniel}}]{Haas_etal_2012}
{Haas}, M.~R., {Schaye}, J., \& {Jeeson-Daniel}, A. 2012, \mnras, 419, 2133

\bibitem[{{Hatton} {et~al.}(2003){Hatton}, {Devriendt}, {Ninin}, {Bouchet},
  {Guiderdoni}, \& {Vibert}}]{Hatton2003}
{Hatton}, S., {Devriendt}, J.~E.~G., {Ninin}, S., {et~al.} 2003, \mnras, 343,
  75

\bibitem[{{Ilbert} {et~al.}(2004){Ilbert}, {Tresse}, {Arnouts}, {Zucca},
  {Bardelli}, {Zamorani}, {Adami}, {Cappi}, {Garilli}, {Le F{\`e}vre},
  {Maccagni}, {Meneux}, {Scaramella}, {Scodeggio}, {Vettolani}, \&
  {Zanichelli}}]{ilbert2004}
{Ilbert}, O., {Tresse}, L., {Arnouts}, S., {et~al.} 2004, \mnras, 351, 541

\bibitem[{{Ilbert} {et~al.}(2005){Ilbert}, {Tresse}, {Zucca}, {Bardelli},
  {Arnouts}, {Zamorani}, {Pozzetti}, {Bottini}, {Garilli}, {Le Brun}, {Le
  F{\`e}vre}, {Maccagni}, {Picat}, {Scaramella}, {Scodeggio}, {Vettolani},
  {Zanichelli}, {Adami}, {Arnaboldi}, {Bolzonella}, {Cappi}, {Charlot},
  {Contini}, {Foucaud}, {Franzetti}, {Gavignaud}, {Guzzo}, {Iovino},
  {McCracken}, {Marano}, {Marinoni}, {Mathez}, {Mazure}, {Meneux}, {Merighi},
  {Paltani}, {Pello}, {Pollo}, {Radovich}, {Bondi}, {Bongiorno}, {Busarello},
  {Ciliegi}, {Lamareille}, {Mellier}, {Merluzzi}, {Ripepi}, \&
  {Rizzo}}]{ilbert2005}
{Ilbert}, O., {Tresse}, L., {Zucca}, E., {et~al.} 2005, \aap, 439, 863

\bibitem[{{Iovino} {et~al.}(2005){Iovino}, {McCracken}, {Garilli}, {Foucaud},
  {Le F{\`e}vre}, {Maccagni}, {Saracco}, {Bardelli}, {Busarello}, {Scodeggio},
  {Zanichelli}, {Paioro}, {Bottini}, {Le Brun}, {Picat}, {Scaramella},
  {Tresse}, {Vettolani}, {Adami}, {Arnaboldi}, {Arnouts}, {Bolzonella},
  {Cappi}, {Charlot}, {Ciliegi}, {Contini}, {Franzetti}, {Gavignaud}, {Guzzo},
  {Ilbert}, {Marano}, {Marinoni}, {Mazure}, {Meneux}, {Merighi}, {Paltani},
  {Pell{\`o}}, {Pollo}, {Pozzetti}, {Radovich}, {Zamorani}, {Zucca}, {Bertin},
  {Bondi}, {Bongiorno}, {Cucciati}, {Gregorini}, {Mathez}, {Mellier},
  {Merluzzi}, {Ripepi}, \& {Rizzo}}]{iovino2005}
{Iovino}, A., {McCracken}, H.~J., {Garilli}, B., {et~al.} 2005, \aap, 442, 423

\bibitem[{{Kaiser}(1987)}]{kaiser1987}
{Kaiser}, N. 1987, \mnras, 227, 1

\bibitem[{{Kauffmann} \& {Haehnelt}(2000)}]{kauffmann00_SAM}
{Kauffmann}, G. \& {Haehnelt}, M. 2000, \mnras, 311, 576

\bibitem[{{Kauffmann} {et~al.}(2004){Kauffmann}, {White}, {Heckman},
  {M{\'e}nard}, {Brinchmann}, {Charlot}, {Tremonti}, \&
  {Brinkmann}}]{Kauffmann2004}
{Kauffmann}, G., {White}, S.~D.~M., {Heckman}, T.~M., {et~al.} 2004, \mnras,
  353, 713

\bibitem[{{Kitzbichler} \& {White}(2007)}]{kitzbichler2007}
{Kitzbichler}, M.~G. \& {White}, S.~D.~M. 2007, \mnras, 376, 2

\bibitem[{{Knobel} {et~al.}(2012){Knobel}, {Lilly}, {Iovino}, {Kova{\v c}},
  {Bschorr}, {Presotto}, {Oesch}, {Kampczyk}, {Carollo}, {Contini}, {Kneib},
  {Le Fevre}, {Mainieri}, {Renzini}, {Scodeggio}, {Zamorani}, {Bardelli},
  {Bolzonella}, {Bongiorno}, {Caputi}, {Cucciati}, {de la Torre}, {de Ravel},
  {Franzetti}, {Garilli}, {Lamareille}, {Le Borgne}, {Le Brun}, {Maier},
  {Mignoli}, {Pello}, {Peng}, {Perez Montero}, {Silverman}, {Tanaka}, {Tasca},
  {Tresse}, {Vergani}, {Zucca}, {Barnes}, {Bordoloi}, {Cappi}, {Cimatti},
  {Coppa}, {Koekemoer}, {L{\'o}pez-Sanjuan}, {McCracken}, {Moresco}, {Nair},
  {Pozzetti}, \& {Welikala}}]{knobel2012_groups20K}
{Knobel}, C., {Lilly}, S.~J., {Iovino}, A., {et~al.} 2012, \apj, 753, 121

\bibitem[{{Komatsu} {et~al.}(2011){Komatsu}, {Smith}, {Dunkley}, {Bennett},
  {Gold}, {Hinshaw}, {Jarosik}, {Larson}, {Nolta}, {Page}, {Spergel},
  {Halpern}, {Hill}, {Kogut}, {Limon}, {Meyer}, {Odegard}, {Tucker}, {Weiland},
  {Wollack}, \& {Wright}}]{komatsu2011_WMAP7}
{Komatsu}, E., {Smith}, K.~M., {Dunkley}, J., {et~al.} 2011, \apjs, 192, 18

\bibitem[{{Larson} {et~al.}(1980){Larson}, {Tinsley}, \&
  {Caldwell}}]{larson1980}
{Larson}, R.~B., {Tinsley}, B.~M., \& {Caldwell}, C.~N. 1980, \apj, 237, 692

\bibitem[{{Le F{\`e}vre} {et~al.}(2004){Le F{\`e}vre}, {Mellier}, {McCracken},
  {Foucaud}, {Gwyn}, {Radovich}, {Dantel-Fort}, {Bertin}, {Moreau},
  {Cuillandre}, {Pierre}, {Le Brun}, {Mazure}, \& {Tresse}}]{lefevre2004b}
{Le F{\`e}vre}, O., {Mellier}, Y., {McCracken}, H.~J., {et~al.} 2004, \aap,
  417, 839

\bibitem[{{Le F{\`e}vre} {et~al.}(2005){Le F{\`e}vre}, {Vettolani}, {Garilli},
  {Tresse}, {Bottini}, {Le Brun}, {Maccagni}, {Picat}, {Scaramella},
  {Scodeggio}, {Zanichelli}, {Adami}, {Arnaboldi}, {Arnouts}, {Bardelli},
  {Bolzonella}, {Cappi}, {Charlot}, {Ciliegi}, {Contini}, {Foucaud},
  {Franzetti}, {Gavignaud}, {Guzzo}, {Ilbert}, {Iovino}, {McCracken}, {Marano},
  {Marinoni}, {Mathez}, {Mazure}, {Meneux}, {Merighi}, {Paltani}, {Pell{\`o}},
  {Pollo}, {Pozzetti}, {Radovich}, {Zamorani}, {Zucca}, {Bondi}, {Bongiorno},
  {Busarello}, {Lamareille}, {Mellier}, {Merluzzi}, {Ripepi}, \&
  {Rizzo}}]{lefevre2005a}
{Le F{\`e}vre}, O., {Vettolani}, G., {Garilli}, B., {et~al.} 2005, \aap, 439,
  845

\bibitem[{{Lemson} \& {Virgo Consortium}(2006)}]{lemson2006_database}
{Lemson}, G. \& {Virgo Consortium}, t. 2006, ArXiv:astro-ph/0608019

\bibitem[{{Lilly} {et~al.}(2009){Lilly}, {LeBrun}, {Maier}, {Mainieri},
  {Mignoli}, {Scodeggio}, {Zamorani}, {Carollo}, {Contini}, {Kneib},
  {LeF{\`e}vre}, {Renzini}, {Bardelli}, {Bolzonella}, {Bongiorno}, {Caputi},
  {Coppa}, {Cucciati}, {de la Torre}, {de Ravel}, {Franzetti}, {Garilli},
  {Iovino}, {Kampczyk}, {Kovac}, {Knobel}, {Lamareille}, {LeBorgne}, {Pello},
  {Peng}, {P{\'e}rez-Montero}, {Ricciardelli}, {Silverman}, {Tanaka}, {Tasca},
  {Tresse}, {Vergani}, {Zucca}, {Ilbert}, {Salvato}, {Oesch}, {Abbas},
  {Bottini}, {Capak}, {Cappi}, {Cassata}, {Cimatti}, {Elvis}, {Fumana},
  {Guzzo}, {Hasinger}, {Koekemoer}, {Leauthaud}, {Maccagni}, {Marinoni},
  {McCracken}, {Memeo}, {Meneux}, {Porciani}, {Pozzetti}, {Sanders},
  {Scaramella}, {Scarlata}, {Scoville}, {Shopbell}, \&
  {Taniguchi}}]{lilly2009_zcosmos}
{Lilly}, S.~J., {LeBrun}, V., {Maier}, C., {et~al.} 2009, \apjs, 184, 218

\bibitem[{{Lynden-Bell}(1971)}]{Lynden_Bell_LF_Cplus}
{Lynden-Bell}, D. 1971, \mnras, 155, 95

\bibitem[{{McCracken} {et~al.}(2003){McCracken}, {Radovich}, {Bertin},
  {Mellier}, {Dantel-Fort}, {Le F{\`e}vre}, {Cuillandre}, {Gwyn}, {Foucaud}, \&
  {Zamorani}}]{mccracken2003}
{McCracken}, H.~J., {Radovich}, M., {Bertin}, E., {et~al.} 2003, \aap, 410, 17

\bibitem[{{Mo} \& {White}(1996)}]{mo1996}
{Mo}, H.~J. \& {White}, S.~D.~M. 1996, \mnras, 282, 347

\bibitem[{{Moore} {et~al.}(1996){Moore}, {Katz}, {Lake}, {Dressler}, \&
  {Oemler}}]{moore1996}
{Moore}, B., {Katz}, N., {Lake}, G., {Dressler}, A., \& {Oemler}, A. 1996,
  \nat, 379, 613

\bibitem[{{Muldrew} {et~al.}(2012){Muldrew}, {Croton}, {Skibba}, {Pearce},
  {Ann}, {Baldry}, {Brough}, {Choi}, {Conselice}, {Cowan}, {Gallazzi}, {Gray},
  {Gr{\"u}tzbauch}, {Li}, {Park}, {Pilipenko}, {Podgorzec}, {Robotham},
  {Wilman}, {Yang}, {Zhang}, \& {Zibetti}}]{Muldrew_etal_2012}
{Muldrew}, S.~I., {Croton}, D.~J., {Skibba}, R.~A., {et~al.} 2012, \mnras, 419,
  2670

\bibitem[{{Peebles}(1980)}]{peebles1980_book}
{Peebles}, P.~J.~E. 1980, {The large-scale structure of the universe}, ed.
  {Peebles, P.~J.~E.}

\bibitem[{{Radovich} {et~al.}(2004){Radovich}, {Arnaboldi}, {Ripepi},
  {Massarotti}, {McCracken}, {Mellier}, {Bertin}, {Zamorani}, {Adami},
  {Bardelli}, {Le F{\`e}vre}, {Foucaud}, {Garilli}, {Scaramella}, {Vettolani},
  {Zanichelli}, \& {Zucca}}]{radovich2004}
{Radovich}, M., {Arnaboldi}, M., {Ripepi}, V., {et~al.} 2004, \aap, 417, 51

\bibitem[{{Sandage} {et~al.}(1979){Sandage}, {Tammann}, \&
  {Yahil}}]{sandage79_LF_STY}
{Sandage}, A., {Tammann}, G.~A., \& {Yahil}, A. 1979, \apj, 232, 352

\bibitem[{{Schechter}(1976)}]{schechter1976}
{Schechter}, P. 1976, \apj, 203, 297

\bibitem[{{Schmidt}(1968)}]{schmidt68_Vmax}
{Schmidt}, M. 1968, \apj, 151, 393

\bibitem[{{Springel} {et~al.}(2005){Springel}, {White}, {Jenkins}, {Frenk},
  {Yoshida}, {Gao}, {Navarro}, {Thacker}, {Croton}, {Helly}, {Peacock}, {Cole},
  {Thomas}, {Couchman}, {Evrard}, {Colberg}, \& {Pearce}}]{springel2005_MILL}
{Springel}, V., {White}, S.~D.~M., {Jenkins}, A., {et~al.} 2005, \nat, 435, 629

\bibitem[{{Springel} {et~al.}(2001){Springel}, {White}, {Tormen}, \&
  {Kauffmann}}]{springel01_SAM}
{Springel}, V., {White}, S.~D.~M., {Tormen}, G., \& {Kauffmann}, G. 2001,
  \mnras, 328, 726

\bibitem[{{Stringer} {et~al.}(2009){Stringer}, {Benson}, {Bundy}, {Ellis}, \&
  {Quetin}}]{Stringer_etal_2009}
{Stringer}, M.~J., {Benson}, A.~J., {Bundy}, K., {Ellis}, R.~S., \& {Quetin},
  E.~L. 2009, \mnras, 393, 1127

\bibitem[{{Temporin} {et~al.}(2008){Temporin}, {Iovino}, {Bolzonella},
  {McCracken}, {Scodeggio}, {Garilli}, {Bottini}, {Le Brun}, {Le F{\`e}vre},
  {Maccagni}, {Picat}, {Scaramella}, {Tresse}, {Vettolani}, {Zanichelli},
  {Adami}, {Arnouts}, {Bardelli}, {Cappi}, {Charlot}, {Ciliegi}, {Contini},
  {Cucciati}, {Foucaud}, {Franzetti}, {Gavignaud}, {Guzzo}, {Ilbert}, {Marano},
  {Marinoni}, {Mazure}, {Meneux}, {Merighi}, {Paltani}, {Pell{\`o}}, {Pollo},
  {Pozzetti}, {Radovich}, {Vergani}, {Zamorani}, {Zucca}, {Bondi}, {Bongiorno},
  {Brinchmann}, {de la Torre}, {Lamareille}, {Mellier}, \&
  {Walcher}}]{temporin2008}
{Temporin}, S., {Iovino}, A., {Bolzonella}, M., {et~al.} 2008, \aap, 482, 81

\bibitem[{{Toomre} \& {Toomre}(1972)}]{toomre1972}
{Toomre}, A. \& {Toomre}, J. 1972, \apj, 178, 623

\bibitem[{{Wang} {et~al.}(2008){Wang}, {De Lucia}, {Kitzbichler}, \&
  {White}}]{wang2008}
{Wang}, J., {De Lucia}, G., {Kitzbichler}, M.~G., \& {White}, S.~D.~M. 2008,
  \mnras, 384, 1301

\bibitem[{{Wang} {et~al.}(2007){Wang}, {Li}, {Kauffmann}, \& {De
  Lucia}}]{wang2007_model}
{Wang}, L., {Li}, C., {Kauffmann}, G., \& {De Lucia}, G. 2007, \mnras, 377,
  1419

\bibitem[{{Weinmann} {et~al.}(2010){Weinmann}, {Kauffmann}, {von der Linden},
  \& {De Lucia}}]{Weinmann_etal_2010}
{Weinmann}, S.~M., {Kauffmann}, G., {von der Linden}, A., \& {De Lucia}, G.
  2010, \mnras, 406, 2249

\bibitem[{{Weinmann} {et~al.}(2006){Weinmann}, {van den Bosch}, {Yang}, {Mo},
  {Croton}, \& {Moore}}]{weinman2006b}
{Weinmann}, S.~M., {van den Bosch}, F.~C., {Yang}, X., {et~al.} 2006, \mnras,
  372, 1161

\bibitem[{{Wilman} {et~al.}(2010){Wilman}, {Zibetti}, \&
  {Budav{\'a}ri}}]{wilman2010_multiscale}
{Wilman}, D.~J., {Zibetti}, S., \& {Budav{\'a}ri}, T. 2010, \mnras, 406, 1701

\bibitem[{{Zucca} {et~al.}(2006){Zucca}, {Ilbert}, {Bardelli}, {Tresse},
  {Zamorani}, {Arnouts}, {Pozzetti}, {Bolzonella}, {McCracken}, {Bottini},
  {Garilli}, {Le Brun}, {Le F{\`e}vre}, {Maccagni}, {Picat}, {Scaramella},
  {Scodeggio}, {Vettolani}, {Zanichelli}, {Adami}, {Arnaboldi}, {Cappi},
  {Charlot}, {Ciliegi}, {Contini}, {Foucaud}, {Franzetti}, {Gavignaud},
  {Guzzo}, {Iovino}, {Marano}, {Marinoni}, {Mazure}, {Meneux}, {Merighi},
  {Paltani}, {Pell{\`o}}, {Pollo}, {Radovich}, {Bondi}, {Bongiorno},
  {Busarello}, {Cucciati}, {Gregorini}, {Lamareille}, {Mathez}, {Mellier},
  {Merluzzi}, {Ripepi}, \& {Rizzo}}]{zucca2006_VVDS_LF}
{Zucca}, E., {Ilbert}, O., {Bardelli}, S., {et~al.} 2006, \aap, 455, 879

\end{thebibliography}

\end{document}